\documentclass[a4,useAMS,usenatbib,usegraphicx]{mn2e}
\usepackage{times}
\usepackage{amsmath}
\include{epsf}
\usepackage{float}
\usepackage{bbding}
\usepackage{epsf}
\usepackage{appendix}
\epsfclipon
\title[Phase mixing of popped star clusters]{Phase mixing due to the Galactic potential: steps in the position and velocity distributions of popped star clusters}
\author[G.~N. Candlish, R. Smith, M. Fellhauer, B.~K. Gibson, P. Kroupa, P. Assmann]{G.~N. Candlish$^{1}$\thanks{E-mail:
gcandlish@astro-udec.cl}, R. Smith$^{1}$, M. Fellhauer$^{1}$, B.~K. Gibson$^{2}$, P. Kroupa$^{3}$, P. Assmann$^{1}$\\
$^{1}$Departamento de Astronom\'ia, Universidad de Concepci\'on, Casila 160-C, Concepci\'on, Chile\\
$^{2}$Jeremiah Horrocks Institute, University of Central Lancashire, Preston, PR1 2HE, UK\\
$^{3}$Argelander Institut f\"ur Astronomie, Universit\"at Bonn, Auf dem H\"ugel 71, D-53121 Bonn, Germany}
\begin{document}

\date{Accepted to MNRAS 6th November 2013}

\pagerange{\pageref{firstpage}--\pageref{lastpage}} \pubyear{2013}

\maketitle

\label{firstpage}

\begin{abstract}
As star clusters are expected to form with low star formation efficiencies, the
gas in the cluster is expelled quickly and early in their development: the star cluster ``pops.'' This leads
to an unbound stellar system, evolving in the Galactic potential. Previous N-body simulations have demonstrated the existence of a stepped number density distribution of cluster stars after popping, both in vertical position and vertical velocity, with a passing resemblance to a Christmas tree. Using numerical and analytical methods, we investigate the source of this structure, which arises due to the phase mixing of the out-of-equilibrium stellar system, determined entirely by the background analytic potential. Considering only the vertical motions, we construct a theoretical model to describe the time evolution of the phase space distribution of stars in a Miyamoto-Nagai disk potential and a full Milky-Way type potential comprising bulge, halo and disk components, which is then compared with N-body simulations. Using our theoretical model, we investigate the possible observational signatures and the feasibility of detection.
\end{abstract}


\section{Introduction}
\label{intro}
Studies of the structure of the Milky Way have suggested the existence of two kinematically distinct stellar populations in the disk. One, with a vertical scale height of $\sim 300$~pc and a vertical velocity dispersion of $\sim 17$~km/s, makes up the \emph{thin disk}, while the \emph{thick disk} is composed of another population with a larger scale height $h_z \approx 1$~kpc and vertical velocity dispersion $\sigma_z \approx 35$~km/s, see e.g. \cite{PasettoThin,PasettoThick}. The vertical velocity dispersion of the thin disk is due to a mixing of populations of various ages, with the younger stars apparently having a lower value of $\sigma_z \approx 2-5$~km/s (\citealp{Fuchs}). Other work, however, suggests that the thin/thick disk distinction is not useful, with there being a smooth, monotonic scale-height distribution (\citealp{Bovy2012}). Assuming that the two-layered structure does exist, its formation has been the subject of much investigation, with several proposed mechanisms.

One such proposal, put forward in \cite{Kroupa}, suggests that the thick disk of stars may have formed due to so-called ``popping'' star clusters. It is now known that the majority of stars form in a clustered mode (see e.g. \citealp{Lada}), and that such clusters typically have low star formation efficiencies. The residual gas of the cluster is then quickly expelled (within no more than a few crossing times of the cluster) by the stellar winds of massive stars and/or supernovae, leading to an unbound stellar system. This mechanism may be responsible for the ``infant mortality'' of star clusters (see e.g. \citealp{Lada,Smith2011}), leading to numerous young star clusters (with ages $< 10$~Myr) and relatively few old clusters (with ages $> 20$~Myr). After gas expulsion, the stars of the cluster begin to expand outwards (the ``popping'' of the cluster), moving under the influence of the general tidal field.

In an attempt to verify the efficacy of this effect in producing the thick disk of the Milky Way, \cite{Assmann} used N-body simulations of such popping star clusters. The velocity dispersions and vertical extent above the disk plane of the resulting distribution of stars, after $10$~Gyr of evolution in an analytic Galactic potential, were then examined to test this thick disk formation scenario. In the course of this work, it was noticed that the vertical number density distribution of the stars, particularly in the early stages of the simulations, exhibited a stepped structure, both in vertical position and vertical velocity. This structure was referred to as the ``Christmas tree,'' and a study of the reason for its development was postponed for later investigation. This is the subject of this paper, in which we will search for a theoretical description of the Christmas tree and we will analyse how likely it is to be observable.

Further motivation for this study came from the observed stepped distributions found in \cite{Kaempf}, \cite{Altmann} and \cite{Maintz}. These distributions, however, are constructed in a different manner from the Christmas tree distribution. In these studies the orbits of the stars were calculated, assuming a background potential, and the distributions show the probability of finding stars at some height $z$ above the disk given those orbits. A stepped structure is apparent in the tails of these distributions. This does not imply, however, the presence of a stepped distribution at any particular point in time, but rather implies the presence of preferred orbits in the sample at large heights above the disk. In our case the stepped distribution is apparent at any particular moment in the dynamical evolution of the cluster, and the form of the distribution itself evolves over time. Moreover, the steps in the observed distributions of \cite{Kaempf}, \cite{Altmann} and \cite{Maintz} occur at around $5$~kpc or more above the disk plane (for the RHB stars, higher for the sdB and RR Lyrae stars). The orbits of stars from even the most massive clusters considered in this work, or in \cite{Assmann}, do not venture further than approximately $10$~kpc, and there are steps in the distribution across the whole range of $z$. Therefore, we do not consider these observed distributions to be connected with those that we will discuss in this paper.

As we will demonstrate, the dynamics of popping star clusters does inevitably lead to the formation of stepped number density distributions. Observing such a distribution could provide important insight into the formation history of the Galactic disk and would constitute a test of the thick disk formation model proposed in \cite{Kroupa}. In the light of future surveys such as GAIA, which will provide detailed kinematic information for approximately one billion stars in our Galaxy, it is important to investigate this phenomenon and to consider the prospects for observations.

\section{Background and theory}
\label{bgrndtheory}
We will demonstrate that the underlying mechanism responsible for the formation of the Christmas tree structure is a phenomenon known as \emph{phase mixing}. This mechanism is well-known as a means by which collisionless stellar systems can relax from out-of-equilibirum states to equilibrium configurations (\cite{LB67,Shu,TremaineHenonLB}). The evolution of a collisionless stellar system is determined by the collisionless Boltzmann equation (CBE):
\begin{equation}
\label{Boltz}
\frac{\partial f(\mathbf{x},\mathbf{v},t)}{\partial t} + \mathbf{v} \cdot \frac{\partial f(\mathbf{x},\mathbf{v},t)}{\partial \mathbf{x}} - \frac{\partial \Phi}{\partial \mathbf{x}} \frac{\partial f(\mathbf{x},\mathbf{v},t)}{\partial \mathbf{v}} = 0,
\end{equation}
where $f(\mathbf{x},\mathbf{v},t)$ is the phase space density distribution and $\Phi$ is the gravitational potential. The CBE states that, along a star's orbit, the density distribution in an infinitesimal region of phase space surrounding that star remains constant. Let us consider the stellar orbits in the vertical $z$ direction only. These orbital trajectories are oscillatory motions in $z$, and phase mixing occurs when those oscillations move out-of-phase with each other. This leads to a shearing of the distribution function in phase space, such that its projection in the $z-w$ plane ($w$ being the velocity in the $z$ direction) begins to resemble a spiral shape. As the system evolves, the spiral arms of the sheared distribution function become increasingly filamentary, while the infinitesimal phase space density around any particle remains constant, as demanded by Eq.~\ref{Boltz}. Eventually, it is no longer possible to resolve the structure with phase space elements of finite size. The best one can practically hope to do (considering also that stellar systems do not constitute an infinite resolution sampling of the distribution function) is to work with the so-called coarse-grained distribution function $\bar{f}$, which may be thought of as a finite resolution sampling of the fine-grained distribution function $f$.

In the general case of a self-gravitating system, there is no straightforward evolution equation for $\bar{f}$ similar to Eq.~\ref{Boltz} (for attempts to develop such a formalism see \citealp{Chavanis}). In the seminal work of Lynden-Bell (\citealp{LB67}), methods of statistical mechanics were used to determine the final form of $\bar{f}$, and therefore the end-point equilibrium of the system\footnote{Although not relevant for our study, it should be pointed out that there are some significant limitations with the standard theory of $\bar{f}$; see \cite{IradBell} and \cite{Dehnen}.}. In our case, however, we may neglect the self-gravity of the system due to the density ratio of cluster to galaxy being low (i.e. the Galactic potential dominates). The stars then proceed to orbit according to the background analytic potential\footnote{While the self-generated cluster potential may be sub-dominant, there are still small self-gravity effects in our N-body simulations: we are not using massless tracer particles. The small influence of such effects, however, essentially amounts to noise in our simulations.}. In principle, therefore, the exact time evolution of $\bar{f}$ \emph{is} known in our case: we must simply determine the orbital trajectories of all the particles in the background potential.

For any reasonably realistic choice for the disk potential there is no guarantee that a closed-form solution for the orbital trajectory of a particle exists. As we are primarily interested in how the various particle oscillations become phase mixed (due to differing frequency of oscillation) our strategy will be instead to determine the frequencies of the particle oscillations, either analytically or by numerical means, and to approximate the true phase space trajectories with simple harmonic motion. We will demonstrate that this simplification is not significant for our conclusions.

At first we simplify our task further in order to highlight the behaviour of phase mixing on the popping star cluster by fixing the radial coordinate in the plane of the disk to be the same for all stars. This effectively treats the disk as infinite in extent. We will use a Miyamoto-Nagai disk (\citealp{MiyamotoNagai}), although we briefly discuss in the Appendix similar results using an infinite exponential disk. Even with this simplification, it is not possible to find a closed-form analytic solution describing all orbital frequencies, and we will instead evaluate some integrals numerically. To gain insight into the parameter dependencies, however, we use analytic asymptotic solutions valid close to and far from the disk. We will refer to our model as the \emph{theoretical} model (either using analytic or numerical results), to distinguish it from the full N-body simulation, with which we will compare. The analysis is then extended by using a full Milky-Way type potential with bulge, halo and disk components in our theoretical model. Again, we will consider only the vertical motions at a fixed radius in the disk. This will also be compared with a full three-dimensional N-body treatment of a cluster moving on a circular orbit around the Galactic centre.

To begin with, in Section~\ref{thesims} we briefly describe the analytic potentials and initial Plummer distributions used in this study. Furthermore we briefly summarise the code used for the comparison N-body simulations. In Section~\ref{reprod} we reproduce the Christmas tree structure in an idealised setting, using a Miyamoto-Nagai potential for a fixed radius in the plane, with the centre of mass of the cluster fixed (i.e. the cluster is not in orbital motion). We then reproduce the Christmas tree in the more realistic scenario of a cluster orbiting in a Galactic potential, with a halo, bulge and disk component. In Section~\ref{theory} we develop our theoretical model for the phase space spiral structure, for the Miyamoto-Nagai disk potential and the full Galactic potential. Using these models we then examine various properties of the ``Christmas tree'' structures, in particular the lifetime of the effect, and the dependence on the parameters of the analytic potential. We also compare our theoretical model against N-body simulations for both the idealised cases considered in Section~\ref{reprod} and more realistic examples using an orbiting cluster. In Section~\ref{obs} we investigate the observability of this phenomenon, considering processes that complicate the picture, such as background field stars, disk heating and low number statistics. We conclude and discuss our results in Section~\ref{concl}.

\section{The N-body simulations}
\label{thesims}
All of our N-body simulations are performed using the particle-mesh code \textsc{Superbox} (\citealp{fellhauer2000}), with equal-mass particles. Such codes neglect the close encounters of stars and are therefore, by definition, collisionless. Furthermore, the number of particles used will often be substantially larger than the typical number of stars in a cluster, allowing us to more accurately sample the phase space distribution of the system. The cluster is always modelled using a Plummer sphere distribution, with a scale length of $1$~pc and a cut off at $20$~pc. This is physically reasonable given that there is no trend of cluster radius with mass (at least for lower mass clusters, \citealp{Murray}). These parameters are, however, relatively unimportant for our study, and while modifying these choices will change the initial distribution of stars somewhat, it makes no difference to our conclusions. More important is the cluster velocity dispersion, which determines the size of the phase space spiral. Therefore we consider three different choices for the initial mass of the clusters: a low mass cluster of $10^5 M_{\odot}$, a medium mass cluster of $10^6 M_{\odot}$ and a high mass cluster of $10^7 M_{\odot}$. The popping of the clusters is simulated by reducing the cluster mass by $80\%$ at the beginning of the simulation, before the particle trajectories are integrated. This is equivalent to a star formation efficiency of $20\%$, with instantaneous gas loss. We consider only instantaneous gas loss for simplicity, but we return to this point in the discussion. Due to the rapid loss of a substantial amount of mass, the cluster is completely unbound from the beginning of the simulation, and therefore the dynamics of the stars are entirely determined by the background analytic potential.

The \textsc{Superbox} code uses several nested grids to ensure that high-resolution grids are placed in regions of interest. In our simulations the self-gravity of the particles is effectively neglected, so we work with rather coarse resolution grids (on the order of $3$~pc for the highest resolution, with $66$~pc and $400$~pc gridboxes for the intermediate and low resolution grids). We have used several choices for both the grid resolution and the timestep size, with no difference to our results.

The analytic disk potentials used in this study are the Miyamoto-Nagai disk, given by
\begin{equation}
\Phi_{\text{MN}}=-\frac{GM_d}{\sqrt{R^2+(a+\sqrt{z^2+b^2})^2}},
\label{MNpot}
\end{equation}
with $M_d = 10^{11} M_{\odot}$, $b = 0.26$~kpc, $a = 6.5$~kpc and $R^2 = x^2 + y^2$. If we fix the radius $R$ we can reduce this potential to be one-dimensional (in effect, the disk becomes infinite in radial extent). The parameter values used here for the Miyamoto-Nagai disk match those used in \cite{Assmann}. The full Milky Way potential may be approximated by superposing a bulge and halo component upon the Miyamoto-Nagai disk. We will use the full potential when we consider the orbital motion of the cluster. We will incorporate the bulge and halo components in our theoretical model, with the restriction of a fixed radius $R$ in the plane of the disk. The bulge is modelled using a Hernquist sphere:
\begin{equation}
\Phi_{\text{Bulge}} = -\frac{GM_b}{r + r_b},
\end{equation}
with $r_b = 0.7$~kpc and $M_b = 3.4 \times 10^{10} M_{\odot}$. The halo component is chosen to be a logarithmic halo given by
\begin{equation}
\Phi_{\text{Halo}} = \frac{v_0^2}{2} \ln(r^2 + d^2),
\end{equation}
with $d = 12$~kpc, $v_0 = 186$~km/s and $r^2 = x^2 + y^2 + z^2$. These parameter choices were originally made in \cite{Johnston95} and have subsequently been used in several studies, e.g. \cite{dinescu, sakamoto, fellhauer2008}. We will also use the simple harmonic oscillator potential to compare with a case in which phase mixing does not arise. This potential is
\begin{equation}
\label{SHOpot}
\Phi_{\text{SHO}} = -\frac{1}{2}\omega_0^2 z^2,
\end{equation}
where $\omega_0$ is the frequency of the simple harmonic oscillator. We will choose this to match the frequency of oscillation of stars close to the plane of the disk in the Miyamoto-Nagai potential.

While the Miyamoto-Nagai disk is known to be a rather poor fit to the true density distribution of the Galactic disk, we consider it here as it is often used in numerical simulations, being a simple three-dimensional potential field of a flattened disk-like system. A brief description of our analysis using an exponential disk is given in the Appendix, but the important results are similar to those using the Miyamoto-Nagai disk.

\section{Reproducing the Christmas tree}
\label{reprod}
The stepped number density distribution described in \cite{Assmann} is due to the phase mixing of the vertical oscillations of the stars in the popping cluster. We will now demonstrate this explicitly with idealised N-body simulations. It is instructive, however, to consider the case of a simple harmonic oscillator potential to show how the stars evolve in phase space when no phase mixing is present.

\subsection{Simple harmonic oscillator potential}
\label{SimSHM}
\begin{figure}
\centering
\includegraphics[width=8.0cm]{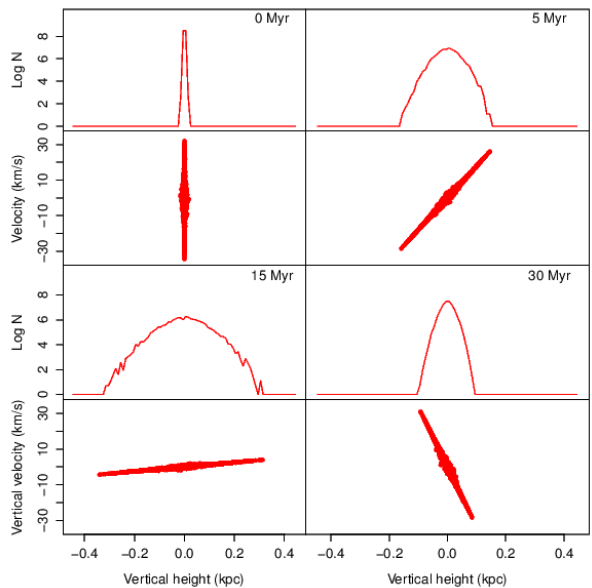}
\caption{Phase space plots of vertical velocities and heights of $10^4$ N-body particles from a $10^5 M_{\odot}$ popping star cluster in a background simple harmonic analytic potential. The log number density in vertical height is also shown. The time of each snapshot is indicated in the upper right corner in each panel. There is clearly no phase mixing in this potential, just endless oscillation.}
\label{SHOphase}
\end{figure}

In a simple harmonic oscillator potential all vertical orbits remain in phase with each other for all time. Therefore, phase mixing cannot occur in this case. In Fig.~\ref{SHOphase} we have shown four snapshots of a $10^5 M_{\odot}$ cluster popping in the background potential given by Eq.~\ref{SHOpot}. These plots show the $z-w$ slice in phase space, where $z$ is the vertical height and $w$ is the vertical velocity. We will refer to this as ``vertical phase space.'' Alongside this, we have plotted the logarithm of the vertical number density to demonstrate the lack of a stepped structure.

The particles are initially localised in $z$, with a spread of vertical velocities. As the cluster expands, the narrow strip of particles rotates around in phase space, with all particles moving in phase with each other (i.e. all particles have the same frequencies of vertical oscillation). Eventually the particles reach the turning points of their orbits, corresponding to the strip of particles lying parallel to the $z$ axis in the plots in Fig.~\ref{SHOphase}. They then collapse back towards the $z=0$ plane, where they again attain their maximum velocities. At this point the strip of particles in phase space has rotated around to lie parallel with the $w$ axis. The collisionless particles then proceed to expand outwards again, until they reach their turning points on the opposite side of the $z=0$ plane (the strip of particles again lies parallel to the $z$ axis). Upon re-collapse to the $z=0$ plane, the strip of particles has completed one full rotation in the vertical phase space. This rotation in phase space (or oscillation in configuration space) continues indefinitely, due to the lack of any relaxation effects (i.e. the potential is static, and there are no two-body encounters).

\subsection{Miyamoto-Nagai disk potential}
\label{SimMN}
The simple harmonic oscillator provides an example of a potential where the popping cluster never phase mixes to some equilibirum distribution. The Miyamoto-Nagai disk, however, exhibits phase mixing. To eliminate the effects of tidal streaming from an orbiting cluster, we simplify the system by placing the centre of mass of the cluster at the origin of the computational domain, without any orbital motion. The analytic potential used is that given in Eq.~\ref{MNpot}, where we set $R=8.5$~kpc, the Galactocentric distance of the Sun. Therefore $R$ plays the role of a parameter and has no relation to the positions of the particles.

\begin{figure}
\centering
\includegraphics[width=8.0cm]{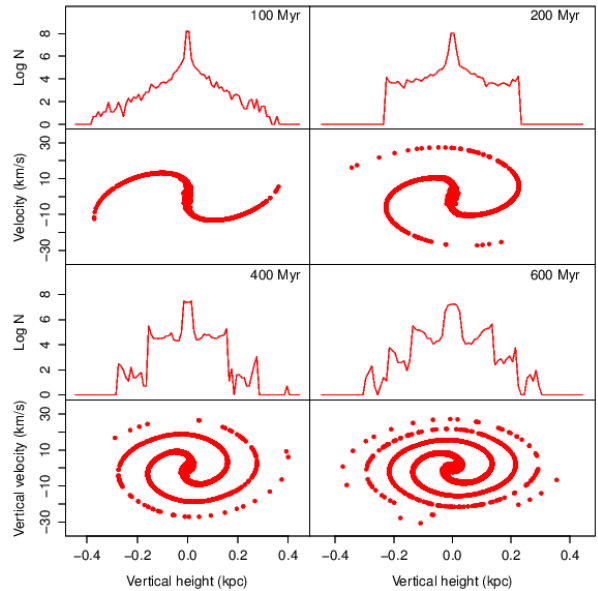}
\caption{Phase space spiral and log vertical number density using $10^4$ N-body particles from a $10^5 M_{\odot}$ popping star cluster in a background Miyamoto-Nagai disk potential, with parameters as specified in Section~\ref{thesims}.}
\label{MNtree}
\end{figure}

In Fig.~\ref{MNtree} we show the particle locations in the vertical phase space at various times. One can clearly see the development of a spiral structure in phase space, and that the spiral winds up as the simulation progresses. This is the phase mixing phenomenon described in Section \ref{bgrndtheory}, due to the differing oscillatory frequencies of the particles. Those particles with large amplitude vertical motions oscillate at a lower frequency than those with smaller amplitude orbits. They therefore rotate in the phase space plane more slowly than the low amplitude particles. This amplitude-dependent phase space rotation then leads to the particles being arranged in a spiral structure. The vertical number density of particles is also shown for each snapshot. The stepped structure in this distribution is the ``Christmas tree'' referred to in \cite{Assmann}.

Fig.~\ref{MNtree} also clearly demonstrates the source of this structure. In constructing the number density distribution in $z$ for some chosen snapshot of the simulation, it is necessary to allocate particles to bins in the $z$ coordinate. From the point-of-view of the phase space spiral, we can see that those particles lying within a spiral arm that crosses the $z$ axis (i.e those particles located at or near the turning points of their orbits) will all fall within a narrow range of $z$ bins, leading to a large number density at those locations. Conversely, those sections of the spiral arms lying approximately parallel to the $z$ axis (i.e. those particles at or near the plane of the disk, where their vertical velocities are highest) contribute to a wider spread of bins in $z$, and thus not many particles are allocated to any particular bin. Of course, there is a central core of stars with low amplitude orbits contributing near to the disk plane, leading to a large number density near $z=0$. In other words, the branches of the Christmas tree (or the steps in the distribution) are co-located with the sections of the phase space spiral that cross the $z$-axis, with the large ``spikes'' in the number density distribution at the ends of each step due to the spiral arms lying almost parallel with the $w$ axis at those points, thus contributing a large number of stars in that $z$ bin.

\begin{figure}
\centering
\includegraphics[width=8.0cm]{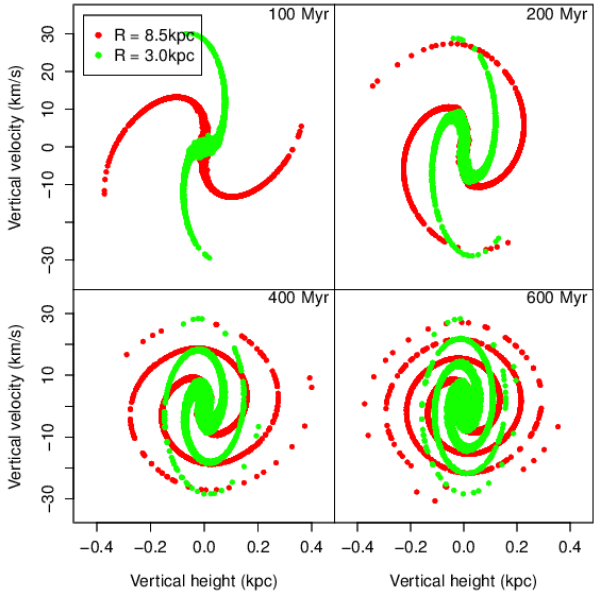}
\caption{Phase space spirals using $10^4$ N-body particles from two $10^5 M_{\odot}$ popping star cluster simulations, both in a background Miyamoto-Nagai disk potential. The red points correspond to the standard $R=8.5$~kpc simulation shown in Fig.~\ref{MNtree}, while the green points correspond to an $R=3.0$~kpc simulation.}
\label{MNtreeR3}
\end{figure}

In Fig.~\ref{MNtreeR3} we show only the phase space spirals of two popping clusters, one where the radial coordinate was fixed at $R=8.5$~kpc, the other where $R=3.0$~kpc. The steeper potential leads to smaller vertical amplitudes for the stellar orbits. Thus the spiral at $R=3$~kpc is less extended in vertical height than the spiral at $R=8.5$~kpc. We can also see that there is a more pronounced central bulge of stars in the cluster at $R=3$~kpc. The increased concentration of stars close to the disk is again due to the steeper potential, leading to more stars in low amplitude orbits. The change in the oscillation frequencies of the particles leads to two distinct effects. The first is that all oscillation frequencies are higher for small $R$, and thus the motions of all the particles in phase space are faster. This is clear from the $100$~Myr panel in Fig.~\ref{MNtreeR3}, where the particles comprising the small $R$ spiral have moved ahead of those in the large $R$ spiral. The other effect is more subtle, and is not obvious from the plot. Changing the value of $R$ in the potential also changes the form of the frequency-amplitude relation of the particles: the relation is steeper at $R=3$~kpc, and so there is a larger difference between the oscillation frequencies for low and high amplitude particles. It is this difference that leads to the spiral structure, and a larger difference means a more tightly wound spiral after a shorter time. We will examine the dependence of the phase space spiral (and therefore the Christmas tree) on the background potential parameters in more detail in Section~\ref{lifetime}.

\subsection{Galactic potential with an orbiting cluster}
\label{SimMW}
The previous examples involved a popping cluster that did not orbit, but was held at a fixed Galactocentric distance. We now consider the more realistic case of an orbiting cluster, in an analytic potential combining a Hernquist bulge, a Miyamoto-Nagai disk and a logarithmic halo, as discussed in Section~\ref{thesims}. The radial coordinates $r$ and $R$ in the background potentials are no longer mere parameters, but are now the usual spherical and cylindrical radii of the particles. We place the centre-of-mass of a $10^5 M_{\odot}$ cluster at $R=8.5$~kpc on a circular orbit with a circular velocity of $\sim 220$~km/s.

\begin{figure}
\centering
\includegraphics[width=8.0cm]{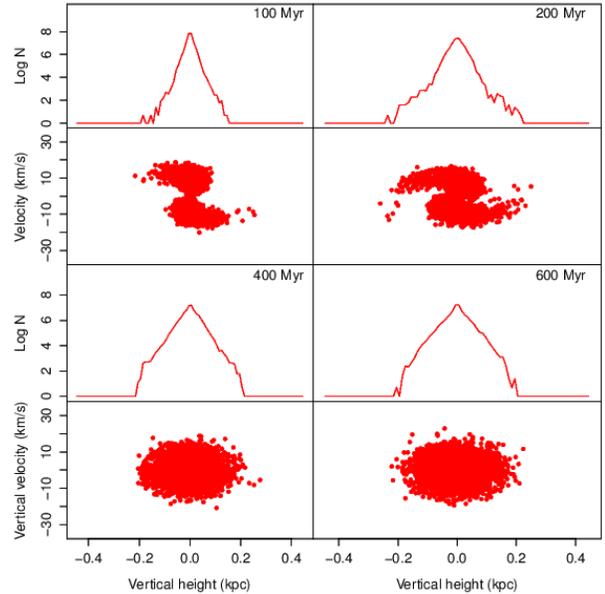}
\caption{Phase space spiral and log vertical number density using $10^4$ N-body particles from a $10^5 M_{\odot}$ popping star cluster orbiting in the full Galactic potential. In this case we have selected stars at random from the full simulation, thus mixing those with purely vertical motions with those on orbits whose radial position changes substantially. The spiral structure is not visible.}
\label{MWalltree}
\end{figure}

The dynamics of the popping cluster in this case is considerably more complex, both due to the orbital motion, and the fact that $R$ is no longer the same for all particles. Stars that execute orbits in which their radial position changes experience a time-varying restoring force from the disk, causing their vertical oscillation frequencies to change with time. This spoils the coherence of the spiral structure. The phase space and vertical number density plots for a random selection from \emph{all} the stars in our orbiting cluster are shown in Fig.~\ref{MWalltree}. There is clearly no spiral structure evident, and therefore no Christmas tree in the vertical number density distribution.

\begin{figure}
\centering
\includegraphics[width=8.0cm]{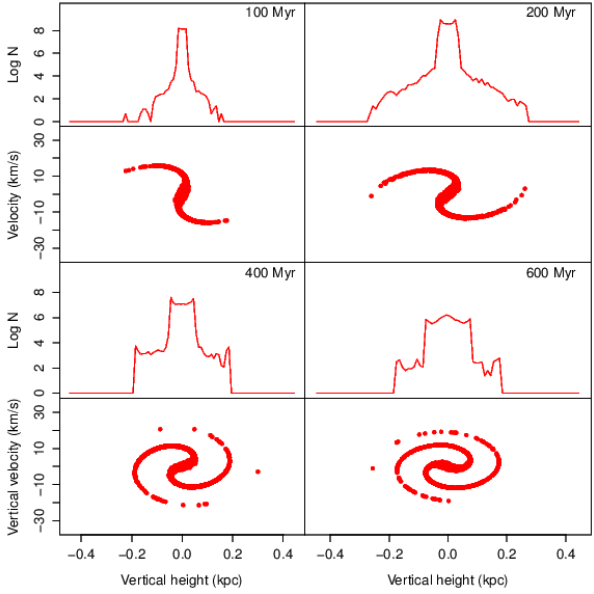}
\caption{Phase space spiral and log vertical number density using various numbers of N-body particles from a $10^5 M_{\odot}$ popping star cluster, where the stars have been selected to be within $50$~pc of the cluster centre-of-mass in the horizontal plane. Selecting such a vertical ``tube'' of stars reveals the spiral structure, as for these stars $R \approx 8.5$~kpc throughout the simulation.}
\label{MWtubetree}
\end{figure}

It is therefore not helpful to consider all the stars in the case of an orbiting cluster. Instead, we will select only those stars whose motion is almost entirely vertical. These stars will move with the centre-of-mass of the cluster around its orbit, and will exhibit the kind of phase space spiral structure discussed earlier. This is akin to selecting a narrow vertical ``tube'' of stars, with the tube centred on the cluster centre of mass. We choose a distance of $50$~pc from the cluster centre-of-mass in the $x$ and $y$ directions. We have done this to produce Fig.~\ref{MWtubetree}, where the spiral structure is again clearly evident. This is because the radial position of these stars in the disk plane does not change significantly as the cluster orbits.

We will see later that the vertical dynamics for an orbiting cluster are most strongly influenced by the disk potential, with the halo and bulge contributing little to the evolution of the spiral structure, except for large amplitude orbits. Provided we select the vertical ``tube'' of stars, the analysis of an orbiting cluster proceeds in a similar manner to that for a cluster that pops in a disk-only potential at fixed radius.

\section{Theoretical model}
\label{theory}
We now proceed to develop a model of the spiral structure that will allow us to investigate various aspects without the need to run full N-body simulations. We will refer to this model as the ``theoretical model'' to distinguish it from the N-body treatment. 

The so-called ``Christmas tree'' number density distribution has been shown to arise due to phase mixing, which causes the particles to arrange themselves into a spiral structure in phase space. As discussed in Section \ref{bgrndtheory}, if the oscillatory vertical motions of the particles have different frequencies, depending on the vertical amplitudes of the orbits (i.e. the $z$ value of the turn-around points) then the particles go out of phase with each other and phase mixing occurs. Therefore to characterise the phase space spiral we must determine the vertical oscillation frequency of each particle trajectory. We do this by calculating the free-fall times for the particles moving in the background analytic Miyamoto-Nagai disk potential and the full Galactic potential of halo, bulge and disk.

\subsection{Free-fall time}
We wish to find the frequency of the particle oscillations in the $z$ direction analytically. In order to do this, we can calculate the time for a particle at $z=A$ to fall to the plane of the disk at $z=0$. From energy conservation, and the fact that $w=0$ at $z=A$ we immediately have
\begin{equation}
\frac{1}{2}w^2 = \Phi(A) - \Phi(z).
\end{equation}
Using $w=dz/dt$ we then have
\begin{equation}
\frac{dz}{dt} = \sqrt{2 (\Phi(A) - \Phi(z))},
\end{equation}
which may be rearranged and integrated to find
\begin{equation}
\label{fftime}
t_{ff} = \int_0^A \left(2 (\Phi(A) - \Phi(z)) \right)^{-1/2} dz,
\end{equation}
where $t_{ff}$ is the free-fall time. For the case of the simple harmonic oscillator potential of Eq.~\ref{SHOpot} this integral is easily computed using the substitution $x \equiv z/A$:
\begin{equation}
t_{ff} = \omega_0^{-1} \int_0^1 (1-x^2)^{-1/2} dx = \omega_0^{-1} \frac{\pi}{2}.
\end{equation}
Taking note of the fact that the free-fall time is related to the full orbital period by $T=4t_{ff}$ (i.e. the free-fall time covers one quarter of the full vertical trajectory), we can find the frequency of the orbit as follows:
\begin{equation}
\label{freqeq}
\Omega = \frac{\pi}{2t_{ff}} = \omega_0,
\end{equation}
as expected for the simple harmonic oscillator.

\subsection{Miyamoto-Nagai disk potential}
We can now apply the same formalism to the Miyamoto-Nagai potential. As we cannot analytically evaluate Eq.~\ref{fftime} for all orbital amplitudes, we will instead focus on the two asymptotic regimes of low amplitude orbits with turning points very close to the plane of the disk and high amplitude orbits with turning points far from the disk. First, we consider the low amplitude orbits, and then the high amplitude orbits. Finally we appeal to a numerical treatment to determine the frequencies at all amplitudes.

\subsubsection{Low amplitude orbits}
\label{lowamporbits}
The Miyamoto-Nagai potential given in Eq.~\ref{MNpot} may be expanded for $z/b \ll 1$ to give
\begin{equation}
\label{MNexpansion}
\Phi_{MN} = -\alpha_0 + \frac{1}{2} \alpha_1 z^2 - \frac{1}{4} \alpha_2 z^4 + \mathcal{O}(z^6),
\end{equation}
where the coefficients of the expansion $\alpha_i$ are functions of the parameters $a,b$, and the planar radius $R$. The explicit expressions for the first three expansion coefficients are given in the Appendix. Substituting this form for the potential in Eq.~\ref{fftime} and taking a further expansion in the dimensionless parameter $\lambda = \alpha_2 A^2/\alpha_1 \ll 1$ we find
\begin{equation}
t_{ff} = \frac{\pi}{2\sqrt{\alpha_1}} + \frac{3\pi \alpha_2 A^2}{16 \alpha_1^{3/2}} + \mathcal{O}(\lambda^2).
\end{equation}
Finally, substituting this into Eq.~\ref{freqeq} and treating $\lambda$ as a small parameter again we obtain for the frequency
\begin{equation}
\label{freqMNlowA}
\Omega = \alpha_1^{1/2} \left( 1 - \frac{3\alpha_2 A^2}{8\alpha_1} + \mathcal{O}(\lambda^2) \right).
\end{equation}
This result may be obtained using perturbation theory to directly solve the equation of motion using the potential in Eq.~\ref{MNexpansion}, which is that of an anharmonic oscillator. For our purposes the important result is the dependence of the frequency on the amplitude of the orbit. It is this behaviour that leads to the phase mixing effect and, ultimately, the ``Christmas tree'' shape of the particle number density distribution. Note that the Miyamoto-Nagai potential behaves as a simple harmonic oscillator very close to the disk, with frequency given by $\alpha_1^{1/2}$.

The expansions performed in the previous Section are only valid for $z \ll b$ and $\alpha_2 A^2/\alpha_1 \ll 1$. Using the parameter choices for the Miyamoto-Nagai disk given in Section~\ref{SimMN} we find that the former constraint is more restrictive, and so the amplitudes of the orbits must satisfy $A \ll 0.26$~kpc for fixed values of the other parameters.

\subsubsection{High amplitude orbits}
Far from the disk the Miyamoto-Nagai potential behaves like that of a point mass, with a $1/z$ dependence. Therefore, for very high amplitude orbits (i.e. $A \gg a,b$ or $R$, where $a$ is the radial scale length of the disk and $R$ is the radial coordinate in the cylindrical polar coordinate system) we have
\begin{equation}
t_{ff} = \frac{\pi}{2} \frac{A^{3/2}}{\sqrt{2GM_d}}.
\end{equation}
This implies an orbital frequency of
\begin{equation}
\label{freqMNhighA}
\Omega = \frac{\sqrt{2GM}}{A^{3/2}}.
\end{equation}
A star executing an orbit with such a large amplitude is, of course, moving well beyond the region of the disk and into the stellar halo. For example, with the disk parameters given in Section~\ref{thesims}, the above orbital frequency is only valid for stars with vertical amplitudes well beyond $8.5$~kpc (recall that we are presently fixing the radial coordinate in the disk potential to this value).

\subsubsection{All orbital amplitudes}
To extend the calculation of the free-fall time beyond the asymptotic regimes of low or high amplitude orbits we must numerically evaluate the integral of Eq.~\ref{fftime}. This is done using a standard numerical integration routine in the statistical programming language R. Using the parameters of the potential from Section~\ref{SimMN} we can easily evaluate the integral for a range of choices of orbital amplitude.

\begin{figure}
\centering
\includegraphics[angle=-90,width=8.0cm]{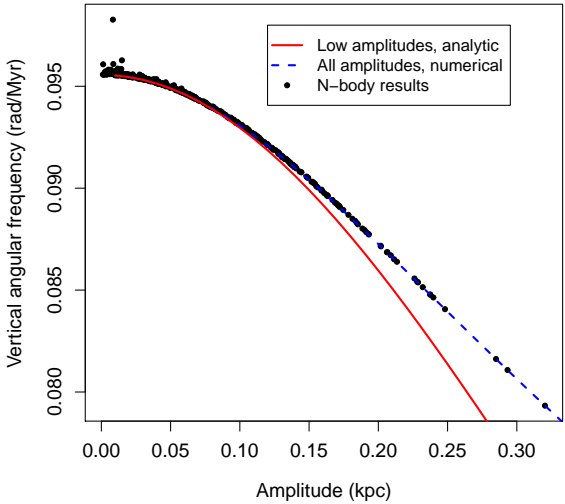}
\caption{Orbital frequencies of 500 N-body particles from a $10^5 M_{\odot}$ cluster in a background Miyamoto-Nagai disk with parameters as in Section~\ref{thesims}. The analytical expression for low amplitude orbits is plotted as the red curve, while the numerical calculation is given by the blue curve.}
\label{MNfftimeSim}
\end{figure}

In Fig.~\ref{MNfftimeSim} we have plotted the vertical oscillatory frequencies of 500 N-body simulation particles (the same simulation shown in Fig.~\ref{MNtree}) as a function of amplitude. These values have been determined from the snapshots of the simulation, by finding a snapshot for which the sign of the particle's vertical velocity is opposite that in the subsequent snapshot. In other words, we find two snapshots such that the particle must have crossed the $z$-axis at some point in the intervening period. This then allows us to interpolate the precise time at which the particle crosses the $z$-axis. After calculating the next time at which the particle crosses the $z$-axis we are then able to subtract these two times to find the half-period of the particle's orbit. These half-periods may then be used to determine the orbital frequency. Superimposed on the particles in Fig.~\ref{MNfftimeSim} is the theoretical prediction from Eq.~\ref{freqMNlowA} (the red line). We see that there is good agreement for the low amplitude orbits where $A \ll b$, but serious divergence from the simulation result for larger amplitude orbits, as expected for our low amplitude approximation. The numerically evaluated relationship between frequency and amplitude (from Eqs.~\ref{fftime} and \ref{freqeq}) is also shown (blue dashed line). This reproduces the simulation result (black points) extremely well for all amplitudes, further demonstrating that self-gravity is playing no significant role. None of the particles in our N-body simulation extend to sufficiently high amplitudes to allow a comparison with Eq.~\ref{freqMNhighA}.

\subsection{Milky-Way type Galactic potential}
\begin{figure}
\centering
\includegraphics[angle=-90,width=8.0cm]{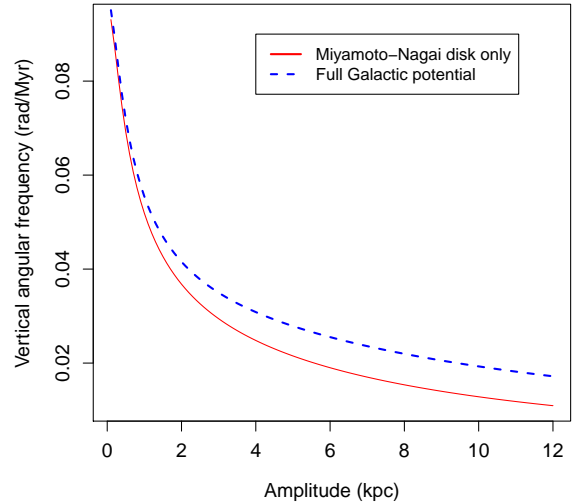}
\caption{Results from numerical calculation of vertical orbital frequencies in both the full Galactic potential and the Miyamoto-Nagai disk only, using the parameters given in Section~\ref{thesims}. The disk dominates the bulge and halo components for small vertical amplitudes, thus the two curves are similar for small $z$. The frequencies are significantly different for large amplitude orbits.}
\label{MWfreqPlot}
\end{figure}

In a similar way, we can numerically determine the vertical orbital frequencies of the stars when a cluster pops in the full Galactic potential. As before, we fix the radial positions of the stars to $R=8.5$~kpc. Performing the calculations for the standard parameters described earlier, including the Hernquist bulge and the logarithmic halo, we find the orbital frequencies shown in Fig.~\ref{MWfreqPlot}. There is good agreement at small amplitudes (i.e. close to the disk where the bulge and halo potentials contribute little to the gravitational potential) between this result and that using only the Miyamoto-Nagai disk, while the large amplitude orbits disagree significantly. This is expected due to the increased contribution of the halo to the gravitational potential further from the disk. The higher frequencies for large amplitude orbits in the presence of the logarithmic halo are due to the increased restoring force. It is also important to note, however, that the \emph{slope} of the frequency-amplitude curve is shallower for the full Galactic potential for large amplitude orbits. This relates to the winding of the spiral, and we shall discuss this more fully in Section~\ref{lifetime}.

\subsection{Determining the amplitudes}
With the preceding theory we can determine the frequency of any particle given its amplitude. Therefore, all that remains is to calculate the amplitude of each particle's orbit, given the initial velocity $w_i$, position $z_i$ and the background analytic potential $\Phi$. This is easily achieved by using conservation of energy and the fact that the particle has no kinetic energy at the turnaround (maximum amplitude, $z=A$) point:
\begin{equation}
\Phi(A) = \Phi(z_i) + \frac{1}{2}w_i^2 = E_i,
\end{equation}
where $E_i$ is the total inital energy (per unit mass). By considering only positive values of $z$ we ensure that the potential $\Phi$ is a monotonic function of $z$, and may be inverted to give
\begin{equation}
A = \Phi^{-1}(E_i).
\end{equation}
The inverse of the Miyamoto-Nagai potential is trivial:
\begin{equation}
A = \left[ \left( \sqrt{(-GM/E_i)^2 - R^2} - a \right)^2 - b^2 \right]^{1/2}.
\end{equation}
The full Galactic potential comprising halo, bulge and disk cannot be inverted to find the amplitude as a function of energy. Therefore we resort to a numerical root-finding procedure, where we solve
\begin{equation}
\Phi_{\text{MN}}(A) + \Phi_{\text{Halo}}(A) + \Phi_{\text{Bulge}}(A) - E_i = 0,
\end{equation}
for the amplitude $A$.

\subsection{Modelling the vertical dynamics}
\label{comp}
We now have all that we need to construct our theoretical model of the popping star cluster evolving in the Miyamoto-Nagai disk potential and the full Galactic potential. Note that we have not determined the true trajectories of the particles. Instead, we have the amplitudes of their orbits and the frequencies of their vertical oscillations. Using this information, and the simplifying assumption that the particle trajectories may be treated as simple harmonic motion, we can approximately reconstruct the phase space evolution of the popping star clusters. Therefore we consider that the vertical position and velocity of any particle is given by
\begin{equation}
\label{eomSHO}
z = A \sin (\Omega(A) t + \phi), \quad \quad w = A \Omega(A) \cos (\Omega(A) t + \phi),
\end{equation}
where $A$ is the orbital amplitude (calculated earlier using energy conservation), $\Omega(A)$ is the frequency (as a function of amplitude, calculated from the free-fall time) and $\phi$ is a phase-shift due to the non-zero initial position of the particle.

Using this approximation of simple harmonic motion for the particle trajectories, with the calculation of the orbital amplitudes and the numerical evaluation of the integral in Eq.~\ref{fftime}, we have arrived at our complete theoretical model. This model is considerably more efficient than a full N-body simulation of the popping cluster, and quickly provides a very good approximation of the phase space spiral at any time we choose.

One may be concerned that the assumption of simple harmonic motion is a simplification that diminishes the usefulness of our model. In fact the assumption is warranted as long as we do not demand that each particle location in phase space in our theoretical model agrees exactly with those in the N-body simulations. The vertical velocities of particles with high amplitude orbits will deviate slightly in our theoretical model from the true values, due to this assumption: we are fixing the amplitudes using energy conservation, and the frequencies are calculated from the analytic potentials, so the $A\Omega(A)$ factor in the expression for $w$ in Eq.~\ref{eomSHO} cannot be adjusted to the actual vertical velocity when the particle crosses the plane of the disk. This means the theoretical spiral will disagree slightly for particles at (or near) $z=0$. The accuracy of our theoretical model will now be demonstrated by comparison with the N-body simulations.

\subsection{Comparing with N-body simulations}
The initial conditions for our theoretical model are taken directly from the initial Plummer distribution of particles used in our N-body simulations. After calculating the orbital amplitudes for all particles we then determine the frequencies associated to these orbits in the manner discussed earlier. We also determine the phase-shifts for the particles due to their non-zero initial vertical positions. Finally, using Eq.~\ref{eomSHO}, we may immediately calculate the (approximate) positions and velocities of each particle at any future time.

\begin{figure}
\centering
\includegraphics[width=8.0cm]{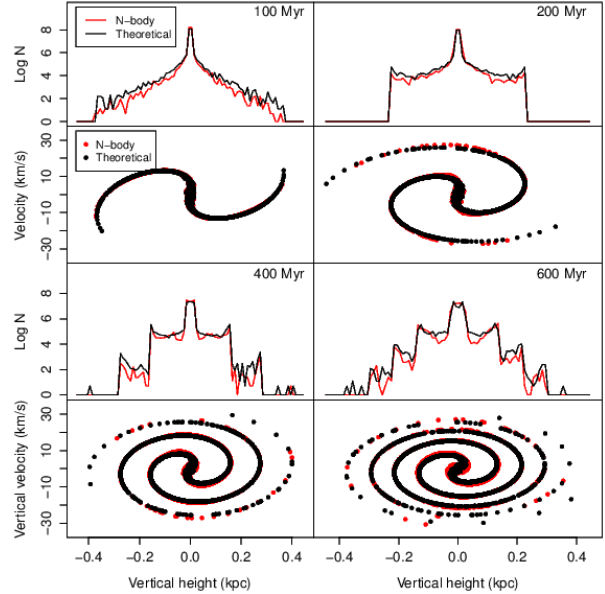}
\caption{Snapshots from the N-body simulation of a $10^5 M_{\odot}$ cluster in the Miyamoto-Nagai disk potential shown in Fig.~\ref{MNtree} with the thoeretical model particles superimposed. The stepped vertical (log) number density distributions are shown, along with the associated phase space spirals. The lines and points in red are for the N-body simulation, while those in black are for the theoretical model.}
\label{analytic_v_sim_MN}
\end{figure}

\begin{figure}
\centering
\includegraphics[width=8.0cm]{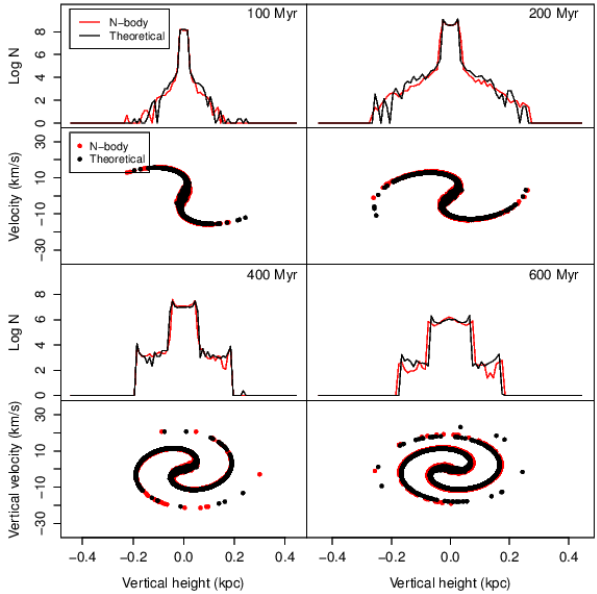}
\caption{Snapshots from the N-body simulation of an orbiting $10^5 M_{\odot}$ cluster in the full Galactic potential shown in Fig.~\ref{MWtubetree} with the thoeretical model particles superimposed. The N-body stars are from the vertical ``tube'' subset discussed earlier. The total number of particles in each N-body simulation snapshot varies due to this selection criteria, while the theoretical model always has $10^4$ particles. Therefore the number density distributions for the theoretical model have been scaled in each snapshot so that the total particle count matches that of the N-body simulation. Note that this simply shifts the distribution up or down, but does not alter the form of the distribution, which matches well with the N-body simulation.}
\label{analytic_v_orbiting}
\end{figure}

In Fig.~\ref{analytic_v_sim_MN} several snapshots of the N-body simulation shown in Fig.~\ref{MNtree} are compared with the theoretical model, and in Fig.~\ref{analytic_v_orbiting} the comparison for the orbiting cluster is shown. To highlight the steps in the number density distribution in these plots we have set the particle count in bins that contain no particles to $1$, to ensure that $\log(n) = 0$ in those bins. We see that there is good agreement between our theoretical model and the N-body simulation.

\begin{figure}
\centering
\includegraphics[width=8.0cm]{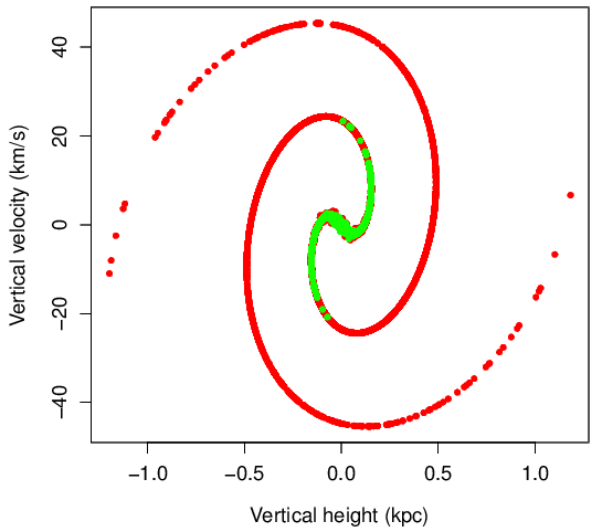}
\caption{Phase space spirals at $150$~Myr after popping for a $10^5 M_{\odot}$ (green) and a $10^6 M_{\odot}$ (red) cluster superimposed. The full Galactic background potential has been used.}
\label{m05vm06spirals}
\end{figure}

The fact that the vertical dynamics of the stars depends only on the background analytic potential is made clear by the excellent agreement between the full N-body simulations and our theoretical model, which only takes the background potential into account. Indeed, we can see in Fig.~\ref{m05vm06spirals} that the spiral structures for a $10^5 M_{\odot}$ and a $10^6 M_{\odot}$ cluster (generated by our theoretical model) are similar, except that the low mass cluster stars do not populate the high velocity/amplitude region accessed by the higher mass cluster. This is simply due to the lower initial energies of the stars in a low mass cluster.

Given that our theoretical model is a very good approximation to the collisionless stellar dynamics in the vertical direction, we now use this model to generate the phase space spirals. Our model is considerably faster to work with than the N-body simulations, and will allow us to quickly investigate, over a wide range of parameter space, the long-term behaviour of the spiral, as well as the dependence of the structure on the parameters of the gravitational potential. We can also use our model to check the dynamics of a cluster that pops away from the plane of the disk.

\subsection{A cluster off the disk plane}
Those stars whose orbits are radially varying experience a time-varying potential, as discussed earlier. If we restrict our attention to a vertical ``tube'' of stars moving on a circular orbit in the plane of the disk, then these stars do not experience a significantly time-varying potential. The excellent agreement between the theoretical model and the N-body simulation for an orbiting cluster supports this assertion.

For a cluster moving on an inclined orbit, with respect to the disk plane, we can again select only those stars in a vertical ``tube'' that coincides with the cluster centre-of-mass. Provided the orbital inclination is not too large, we may again approximate the vertical dynamics using our theoretical model.

\begin{figure}
\centering
\includegraphics[width=8.0cm,angle=-90]{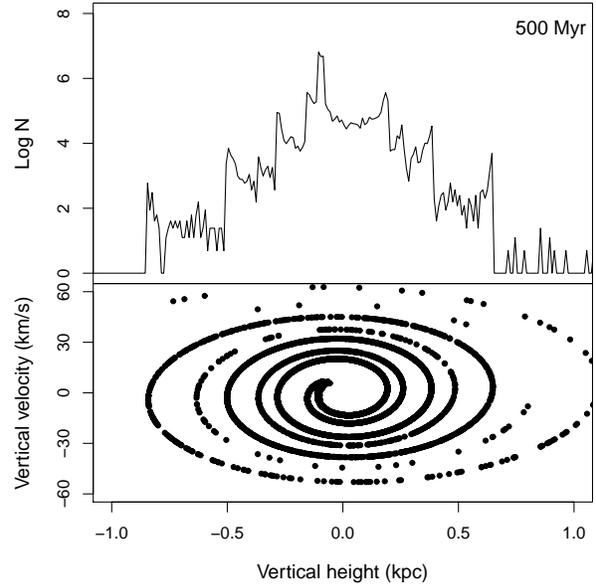}
\caption{Phase space spirals and log number density distributions (in position) for a $10^6 M_{\odot}$ cluster that pops with centre-of-mass at $z=0.1$~kpc and $w=10$~km/s in phase space coordinates.}
\label{offplanepop}
\end{figure}

The popping of the cluster now takes place off the disk plane, leading to modified initial conditions for the stars. The initial vertical positions and velocities are all now shifted with respect to those used earlier by the centre-of-mass vertical position and velocity of the cluster at the time of popping. This leads to an asymmetric spiral, as shown in Fig.\ref{offplanepop}, for a $10^6 M_{\odot}$ cluster that pops with centre-of-mass located at $z=0.1$~kpc and $w=10.0$~km/s in phase space coordinates. Moving the cluster off the disk plane also leads to an asymmetric stepped number density distribution. The basic form of the distribution, however, is the same as before. For the rest of this study we will restrict ourselves to the simpler case of a symmetric phase space spiral, as our main conclusions are not affected by this shifting of the spiral.

\subsection{Lifetime of structure and parameter dependence}
\label{lifetime}
\begin{figure}
\centering
\includegraphics[width=8.0cm,angle=-90]{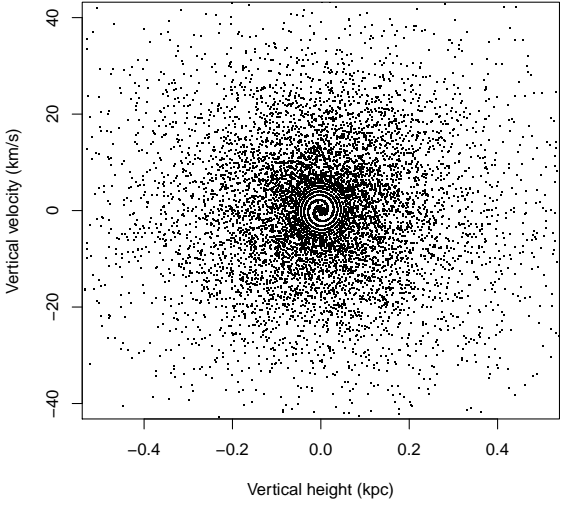}
\caption{Snapshot after $20$ Gyr of evolution of our theoretical model, for a cluster of mass $10^6 M_{\odot}$ with $10^4$ particles. Further from the spiral centre, where the particle density is low, the spiral structure is no longer apparent. It can still be seen in the central region, given the resolution of the plot.}
\label{truelifetime}
\end{figure}

In Fig.~\ref{truelifetime} we show the results of our theoretical model for a medium mass cluster after a very long term evolution ($20$~Gyr) in the full Milky-Way type potential (with the radius fixed at $8.5$~kpc). In the region of phase space where the particle density is low, the spiral structure is no longer visible. In the high density central region the structure persists. One can imagine, however, plotting this figure at ``lower resolution'' by using larger plotting symbols, which would erase the spiral. The point is that the structure still exists, even after such a long term evolution, and may be seen if one is able to observe the fine-grained detail of the phase space distribution, and if one has sufficiently well-sampled the phase space. Note that, for an orbiting cluster, there will be some contamination from interloper stars that are not executing entirely vertical motion (as they orbit with the cluster centre-of-mass), and from stars in the tidal streams as the cluster orbits.

If the spiral structure is always present \emph{in principle}, one may ask how long it would take for the spiral arms to become sufficiently wound up as to be \emph{in practice} indistinguishable. To get a handle on the lifetime of the structure in this sense, we calculate the length of time taken for a certain number of spiral arms to lie within a range of $z$ amplitudes, using the numerically evaluated frequency-amplitude relation. The timescales for 10 spiral arms to lie in the range $0.05 \le z \le 0.1$~kpc for different parameter choices of the Miyamoto-Nagai disk potential are given in Table~\ref{lifetimeTableMN}, while those for the full Galactic potential, are given in Table~\ref{lifetimeTableMW}. The parameters have been varied individually, with the rest kept fixed at the ``standard'' values given in Section~\ref{thesims}. Each parameter is then either doubled or halved in value. Note that 10 spiral arms within a $50$~pc range is not a particularly high ``density'' of spiral arms in phase space: the arms are spaced roughly $5$~pc apart. From both Table~\ref{lifetimeTableMN} and Table~\ref{lifetimeTableMW} we see that the typical timescales for this level of spiral development are very long. The spiral structure takes an extremely long time to become wound up beyond the point where the arms are (in principle) distinguishable.

\begin{table}
\caption{Timescale for the formation of 10 spiral arms in the range $0.05 \le z \le 0.1$~kpc, with various choices of parameters in the Miyamoto-Nagai potential.}
\label{lifetimeTableMN}
\begin{tabular}{ c | c | c | c | c }
\hline
$R$ (kpc) & $a$ (kpc) & $b$ (kpc) & $M_d$ ($M_{\odot}$) & Timescale (Gyr) \\
\hline
$8.5$     & $6.5$     & $0.26$    & $10^{11}$           & 17.2            \\
\hline
$3.25$    & $6.5$     & $0.26$    & $10^{11}$           & 10.4            \\
$17.0$    & $6.5$     & $0.26$    & $10^{11}$           & 38.7            \\
\hline
$8.5$     & $3.25$     & $0.26$    & $10^{11}$          & 19.5            \\
$8.5$     & $13.0$     & $0.26$    & $10^{11}$          & 21.1            \\
\hline
$8.5$     & $6.5$     & $0.13$    & $10^{11}$           & 3.8            \\
$8.5$     & $6.5$     & $0.52$    & $10^{11}$           & 90.4            \\
\hline
$8.5$     & $6.5$     & $0.26$    & $5 \times 10^{10}$  & 24.3            \\
$8.5$     & $6.5$     & $0.26$    & $2 \times 10^{11}$  & 12.2            \\
\hline
\end{tabular}
\end{table}

\begin{table}
\caption{Timescale for the formation of 10 spiral arms in the range $0.05 \le z \le 0.1$~kpc, with various choices of parameters in the full Galactic (halo, bulge and disk) potential. The disk parameters have been set to the ``standard'' values of Section~\ref{thesims}.}
\label{lifetimeTableMW}
\begin{tabular}{ c | c | c | c | c }
\hline
$v_0$ (km/s) & $d$ (kpc) & $r_b$ (kpc) & $M_b$ ($M_{\odot}$)  & Timescale (Gyr) \\
\hline
$186$        & $12.0$    & $0.7$       & $3.4 \times 10^{10}$ & 17.6            \\
\hline
$372$        & $12.0$    & $0.7$       & $3.4 \times 10^{10}$ & 18.0            \\
$93$         & $12.0$    & $0.7$       & $3.4 \times 10^{10}$ & 17.4            \\
\hline
$186$        & $24.0$    & $0.7$       & $3.4 \times 10^{10}$ & 17.5            \\
$186$        & $6.0$     & $0.7$       & $3.4 \times 10^{10}$ & 17.7            \\
\hline
$186$        & $12.0$    & $1.4$       & $3.4 \times 10^{10}$ & 17.5            \\
$186$        & $12.0$    & $0.35$      & $3.4 \times 10^{10}$ & 17.6            \\
\hline
$186$        & $12.0$    & $0.7$       & $6.8 \times 10^{10}$ & 17.8            \\
$186$        & $12.0$    & $0.7$       & $1.7 \times 10^{10}$ & 17.5            \\
\hline
\end{tabular}
\end{table}

The results in Table~\ref{lifetimeTableMN} show that the spiral takes longer to wind up for larger $R$, larger radial scale length $a$ and larger vertical scale height $b$. As expected, the spiral also winds up more slowly for a less massive disk potential. Furthermore, the spiral winding takes longer for a \emph{lower} value of radial scale length $a$, indicating that there is a minimum in the winding timescale for an intermediate value of $a$ (assuming all other parameters are fixed). The vertical scale height parameter $b$ is easily the most sensitive parameter, causing large changes in the winding timescale. Variations of the bulge and halo parameters are far less significant than those of the disk. Note that the spiral winding timescale is slightly longer for the full Galactic potential (with standard parameters) than for the Miyamoto-Nagai disk alone. This is due to the slightly lower slope in the frequency-amplitude relation for the Galactic potential shown in Fig.~\ref{MWfreqPlot}. When a given change in amplitude leads to a \emph{lower} change in frequency, the phase mixing effect is diminished, leading to a longer spiral winding timescale. In Tables~\ref{lifetimeTableMN} and \ref{lifetimeTableMW} we are considering only low amplitude orbits, where the difference in the curves in Fig.~\ref{MWfreqPlot} is much smaller. Therefore the difference in winding timescale is small.

Our reference value for the vertical scale height parameter $b$ has been $0.26$~kpc. Taking $0.2-0.3$~kpc to be a realistic range of values of this parameter to model the disk of our Galaxy, we find that the timescale for 10 phase space spiral arms to form in the range $0.05 \le z \le 0.1$~kpc is between $9.5$ and $24.0$~Gyr. Therefore, for all other parameters being fixed, the timescale for the spiral winding can vary significantly within a small range of realistic vertical scale heights.

Why does the spiral winding depend on these parameters? The spiral structure forms due to the contribution of anharmonic terms to the restoring force that acts on the stars, as discussed earlier. The particles in a simple harmonic oscillator potential never phase mix, by construction, and so increasing the restoring force in this case simply leads to more rapid in-phase oscillations: the ``spiral winding timescale'' in this case would be infinity. It is therefore the contribution of the anharmonic terms that affect the spiral winding timescale. Equivalently, the degree of slope in the frequency-amplitude indicates the influence of the anharmonic terms in deviating the behaviour from harmonic oscillation. If a parameter choice boosts the contribution of the anharmonic behaviour then the spiral structure will wind up more rapidly.

Using our analytic result for low amplitude orbits in the Miyamoto-Nagai disk, Eq.~\ref{freqMNlowA}, we can investigate the contribution of the anharmonic term given changes to each of the parameters in the disk potential. We can keep two of the parameters fixed while we vary the other, in order to determine the effect of that parameter on the value of the anharmonic term. When we fix the other parameters we choose the values $a=6.5$~kpc, $b=0.26$~kpc and $R=8.5$~kpc as before, and we fix the orbital amplitude at $A=0.1$kpc. Furthermore, the vertical scale length $b$ is not varied below the value $b=0.2$~kpc, as the original series expansion is only valid for $z \ll b$ anyway. With these choices, we have ensured that the sub-leading term in Eq.~\ref{freqMNlowA} is always much smaller than the leading order harmonic term for all parameter values and $A$ is less than the vertical scale length of the disk $b$, maintaining the validity of the series expansion. 

\begin{figure}
\centering
\includegraphics[angle=-90,width=8.0cm]{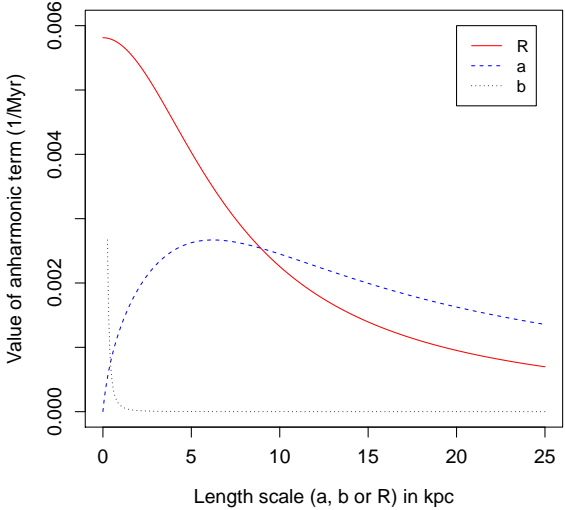}
\caption{Effect of varying the radial and vertical scale parameters of the Miyamoto-Nagai disk, or the radius at which the cluster is placed, on the value of the anharmonic term in Eq.~\ref{freqMNlowA}, the low amplitude equation for the vertical oscillation frequency.}
\label{params_anharm}
\end{figure}

In Fig.~\ref{params_anharm} we show the dependence of the sub-leading term on various choices for the parameter values. For the vertical scale height $b$ it is clear that the anharmonic term becomes a negligible contribution to the frequency for larger scale heights, and the transition to such a negligible contribution is rapid, reflecting the sensitivity of the winding timescale on this parameter shown in Table~\ref{lifetimeTableMN}. Conversely, for small scale heights the anharmonic term becomes increasingly important, although we must be cautious as our series expansion is not valid for excessively small values of $b$. For the radial scale parameter $a$ the situation is reversed: a small $a$ puts us in the harmonic regime, while a larger $a$ leads to an increased anharmonic contribution, which begins to decrease again for very large values. Finally, if the cluster is further out in the disk, i.e. with larger $R$, then the anharmonic contribution decreases. Thus all the behaviour exhibited in Table~\ref{lifetimeTableMN} can be attributed to the contribution of anharmonic behaviour to the vertical stellar dynamics.

\section{Observability}
\label{obs}
Using our theoretical model, we now investigate the feasibility of observing the spiral structure (or stepped number density distribution). Firstly, we look at the effect of low number statistics, showing that a small number of steps is easier to observe than a large number. Then we consider the timescales involved for such a step to develop. Next we add contamination from a background star field, which may be so noisy as to overwhelm the possibility of detection of any steps in the number density distribution. Over time the structure is susceptible to the effects of kinematic heating in the Galactic disk, which may wash out any observable features. Therefore we investigate the effects on the phase space spiral of randomising the stellar velocities. Finally, we check to see if typical observational errors also render the structure unobservable.

\subsection{Low resolution sampling}
\label{lownstats}
\begin{figure}
\centering
\includegraphics[angle=-90,width=8.0cm]{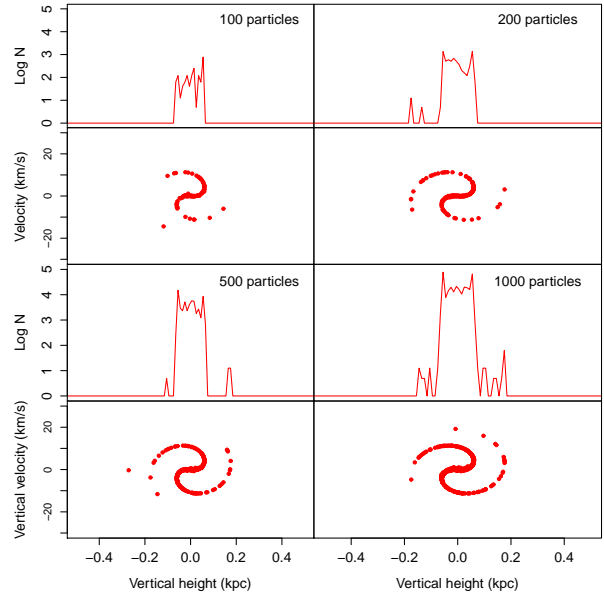}
\caption{The effect of choosing only a small subset of particles in the $10^5 M_{\odot}$ cluster. The full Galactic background potential has been used. The cluster has evolved for $500$~Myr after popping.}
\label{lowNlowmassMW}
\end{figure}

\begin{figure}
\centering
\includegraphics[angle=-90,width=8.0cm]{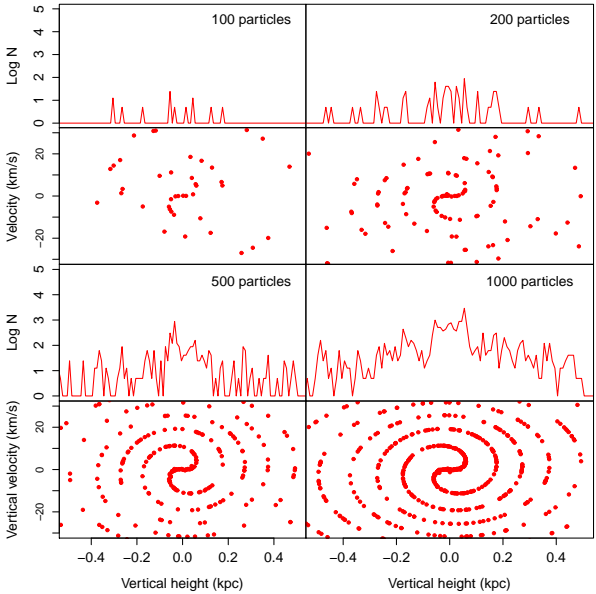}
\caption{The effect of choosing only a small subset of particles in the $10^7 M_{\odot}$ cluster. The full Galactic background potential has been used. The cluster has evolved for $500$~Myr after popping.}
\label{lowNhighmassMW}
\end{figure}

A major problem for detection of the phase space structure is the finite number of stars being used to sample the phase space distribution. We have already seen from Fig.~\ref{truelifetime} that the spiral is not visible if the phase space is not sufficiently well sampled. We will now investigate the observational consequences of this, using random subsamples of our full set of star particles. The results for the full Galactic potential, with a low mass ($10^5 M_{\odot}$) and high mass ($10^7 M_{\odot}$) cluster, are shown in Figs.~\ref{lowNlowmassMW} and \ref{lowNhighmassMW}. For now we arbitrarily choose a time of $500$~Myr after popping to observe the stepped number density distribution. Later we include background stars, but for the moment this may be considered the observationally ideal case when the background stars have been perfectly subtracted away from the cluster stars.

Comparing Figs.~\ref{lowNlowmassMW} and \ref{lowNhighmassMW}, it is clear that all of the stars of the low mass cluster are concentrated in a small region of $z$, where there are two steps in the density distribution. Therefore, as we add more stars to our subsample of the cluster, we sample these two steps more effectively, and they become evident almost immediately. For the high mass cluster, however, the stars have much higher velocity and position dispersions, leading to a larger spiral with more steps in the distribution. Now the random subsample is more likely to include stars far out from the central regions of the spiral. The structure is then much noisier for low numbers of stars, and from the plot it is clear that the steps in the distribution are not particularly evident even for a sample of 1000 stars. If our sample of cluster stars is, instead, attempting to ``fill in'' only a small number of steps, then we have more chance of detecting them, especially when there is background noise.

One can conclude from this that the ideal circumstances for detection are to have many cluster stars and a small number of steps. The first criterion implies that a high mass cluster is, in principle, more detectable. As the number of steps increases with time, the second criterion implies a higher detectability early in the evolution of the phase space spiral.

When in the spiral evolution do these steps first arise? As the comparison with the mock observations (see the next Section) is easier using vertical velocities, we will determine the time at which steps develop in the vertical velocity number density. Similar conclusions apply for the vertical positions. It is a somewhat subjective task to decide when the steps have formed. We simply judge by eye, choosing a time when the central ``bar'' lies horizontally in phase space. The full Galactic potential with the parameters given in Section~\ref{thesims} is used. The results are shown in Table~\ref{stepTimeTable}.

\begin{table}
\caption{Timescale for the formation of a small number of steps in the number density distribution (in vertical velocity).}
\label{stepTimeTable}
\begin{tabular}{ c | c | c | c | c }
\hline
Cluster mass $M_{\odot}$ & 1 step (Myr) & 2 steps (Myr) \\
\hline
$10^5$ & $135$ & $420$ \\
$10^6$ & $35$ & $95$ \\       
$10^7$ & $35$ & $65$ \\       
\hline
\end{tabular}
\end{table}

These results again support the contention that it is observationally easier to detect the structure for high mass clusters, as these will form a single large step earlier in their evolution, leaving less time for other effects, such as disk heating, or the mixing from radial motions of the stars, to wash out the spiral structure. We will examine disk heating in Section~\ref{diskheatingsect}.

\subsection{Background contamination}
\label{bgndcont}
The ideal circumstance for detection of the steps in the density distribution is, of course, when we are able to ignore the background field completely. This would be the case if we were able to clearly discriminate the cluster stars from the background field, as would be possible with a clear age distinction between the very young stars of a recently popped cluster and the old background population. As stated earlier, however, even without any background noise we must still sample the phase space sufficiently well to be able detect the structure.

Let us now complicate matters by considering noise from a background field of stars. We do this by using a mock star catalogue for the Milky Way (\citealp{MWmodel}), with an upper limit of $M_V = 20.0$ on the absolute (V-band) magnitude of the observed stars. We have chosen two different directions of view, both parallel to the disk plane, one looking towards the bulge (i.e. Galactic latitude and longitude of $l \approx 0, b \approx 0$) and the other looking away from the bulge ($l \approx 180, b \approx 0$). Depending on the direction of view, the number counts of background stars generated by the model vary significantly. We then use this background field to determine how much ``noise'' will contaminate the signal of the cluster number density distribution.

Our strategy is as follows: we fit a smooth distribution to the background stars, which is then subtracted to estimate the noise associated from the background. The absolute value of the residual is used, as we wish to generate noise counts in each velocity bin. The standard deviation and mean of the residual's absolute value is then used to generate Gaussian noise in all vertical velocity bins. The velocity bins have a width of $2$~km/s.

We use our theoretical model (with the full Galactic potential) to generate a cluster number density distribution in vertical velocity space, with some subsample $N$ of cluster stars. A $10^7 M_{\odot}$ cluster is used, $75$~Myr after popping, when two steps have formed in the number density distribution, one at approximately $\pm 14$~km/s and another at approximately $\pm 74$~km/s. The ``star counts'' from our noise are added up for all velocity bins in which the cluster signal appears. This value is then divided into $N$ to determine a simple signal-to-noise estimation.

\begin{figure}
\centering
\includegraphics[angle=-90,width=8.0cm]{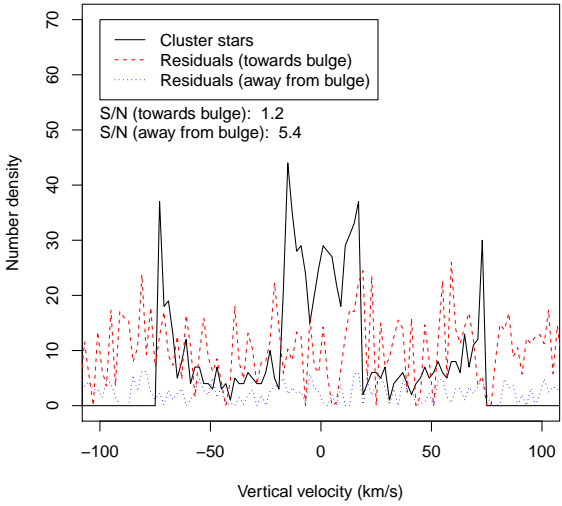}
\caption{Vertical velocity number density distribution for $900$ stars from a $10^7 M_{\odot}$ cluster $65$~Myr after popping. The Gaussian noise values generated using the mock catalogue distribution are also shown (absolute values of the residuals have been used, and thus the noise has a non-zero mean).}
\label{exampleobscluster}
\end{figure}

The results are shown in Fig.~\ref{exampleobscluster} for a cluster subsample of $900$ stars. We see that, when looking towards the bulge, the background noise is comparable to the cluster signal. When looking away from the bulge, however, the background is far less noisy, and thus the cluster steps are much clearer.

We may consider this to be a best-case scenario for detection of the cluster, as we are assuming that we can select those stars that are moving almost entirely vertically after the cluster pops (see Section \ref{SimMW}). It is clear that detection of the cluster will depend strongly on the amount of noise arising from the background, and the number of observed cluster stars. A significantly more noisy background increases the number of cluster stars required for detection of the steps. Assuming that an observation of the cluster stars selects from all possible velocities with uniform probability, then the total number of cluster stars that must be detected increases. Furthermore, for a given number of such stars, we obviously achieve a higher count in each step when there is a small number of steps. This again suggests that, for maximum detectability, we must catch a high mass cluster early in its evolution when fewer steps have formed.

\subsection{Disk heating}
\label{diskheatingsect}
It is well known that various physical mechanisms, such as transient spiral structure or scattering off giant molecular clouds (see e.g. \citealp{Jenkins, Spitzer, Lacey}), act in spiral galaxies to increase the vertical velocity dispersions of the disk stars, referred to as ``disk heating'' (for a recent discussion see \citealp{Gerssen}).  We use our theoretical model to consider the effect of disk heating upon the phase space spiral structure. We will treat this in a very simple manner, by merely adjusting the vertical velocities of the stars by a random amount, drawn from a normal distribution, i.e. $w' = w + \lambda$ where $\lambda$ is a normally distributed random variable with zero mean, so we do not change the centre-of-mass velocity of the cluster. In so doing, we can in fact adjust the vertical velocity dispersion of our particles until it more closely matches that of a mature thick (or thin) disk. Our purpose is not to investigate the role of this mechanism in the formation of the disk, but simply the effect upon the phase space structure. There is some controversy in the literature (for example, \citealp{Carlberg, Seabroke, Soubiran, Quillen} among others) regarding the time dependence of the age-velocity dispersion relation: is there a power-law time dependence, does the velocity dispersion level off at $~5$~Gyr or are the heating events discrete step changes? For simplicity we apply the ``heating'' at specific time steps during the evolution of the phase space structure, with the amount of heating chosen so that the final vertical velocity dispersion is $\sigma_z \approx 40$~km/s, approximately that of the thick disk. After each heating event the amplitudes, frequencies and phase-shifts of the particles are re-calculated, and the model evolves using the simple harmonic motion approximation as before.

\begin{figure}
\centering
\includegraphics[width=8.0cm]{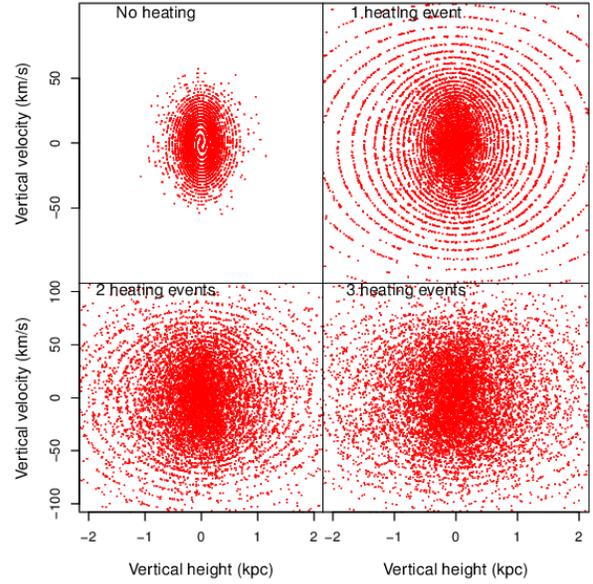}
\caption{The effects of disk heating on the spiral structure of the $10^6 M_{\odot}$ cluster in the full Galactic potential. All plots are $1.3$~Gyr into the cluster evolution.}
\label{diskheating}
\end{figure}

Our simple heating model is applied to the $10^6 M_{\odot}$ cluster (with initial vertical velocity dispersion of $\sigma_z \approx 13$~km/s) with one, two or three heating events. The details of the heating model used are given in Table~\ref{heatingmodel}. As we are demanding a fixed final velocity dispersion, the total amount of kinetic energy added to the system is very similar in each case. The phase space distributions at $1.3$~Gyr after popping are shown in Fig.~\ref{diskheating}. At this time the cluster has already been heated, and each particle has returned to vertical simple harmonic oscillations, albeit with new orbital parameters as discussed above. The timing of the heating events is arbitrary: after the cluster returns to the conventional oscillatory evolution the vertical velocity dispersion remains roughly constant. Our theoretical model, by construction, cannot modify the velocity dispersion of the cluster (the particle trajectories in phase space are closed loops). This is physically sensible if we can neglect self-gravity, which we have shown to have a negligible influence in the N-body simulations.

\begin{table}
\caption{Parameters of the disk heating model used to generate the results in Fig.~\ref{diskheating}. The first column is the number of heating events, the second is the time after popping at which the heating is applied. The third column is the standard deviation of the normally distributed random velocity shift, $\lambda$. The fourth column is the increase in the (rms) kinetic energy, assuming $10^4$ particles in the cluster, and thus a mass-per-particle of $10^2 M_{\odot}$.}
\label{heatingmodel}
\begin{tabular}{ c | c | c | c }
\hline
Number & Times & $\sigma(\lambda)$ (km/s) & $\Delta KE$ ($M_{\odot}$~km$^2$ s$^{-2}$) \\
\hline
1 & $1$~Gyr & $63.0$ & $7.1 \times 10^4$ \\
2 & $1$, $1.1$~Gyr & $46$ & $7.2 \times 10^4$ \\       
3 & $1$, $1.1$, $1.2$~Gyr & $38$ & $7.2 \times 10^4$ \\       
\hline
\end{tabular}
\end{table}

It is clear that the spiral structure is significantly spread in vertical velocity and vertical height by the heating, due to the high vertical velocity dispersion that we seek. After a single heating event, the central regions of the spiral are smoothed out, but some structure remains at higher velocities/amplitudes. As more heating events are applied, however, the particles become more randomised, eventually smoothing out the phase space distribution completely. Therefore, as stated previously, we stand a better chance of detecting the stepped distribution if we can catch a massive cluster, which is less affected by any heating (the higher amplitude regions may still exhibit the phase space spiral), early after popping (as there is less time for heating to disrupt the structure).

\subsection{Observational errors}
The effect of observational errors can be modelled by adding Gaussian random noise to the measurements of the phase space coordinates of the particles.

We will asssume that the accuracy of future stellar surveys allows the measurement of positions and proper motions on the sky to an accuracy of $\sim 200$~$\mu$as (the accuracy depends on the magnitude, with a range from $\sim 25$~$\mu$as for 15th G-magnitude stars down to $\sim 330$~$\mu$as for 20th G-magnitude stars for the case of GAIA, as stated in \citealp{GAIA}). Given this, a cluster that pops at a distance of $\sim 3$~kpc will have tangential velocity measurement errors of around $3$~km/s, while a cluster at the large distance of $\sim 30$~kpc will have tangential velocity errors of $\sim 30$~km/s. With this astrometric accuracy, the transverse position error, even at $30$~kpc, is $\sim 3 \times 10^{-5}$~pc. The distance error, however, is considerably larger, on the order of a few percent (\citealp{GAIA}). Considering a large error of 10\% for a cluster at $3$~kpc (i.e. an error of $300$~pc in distance), this is six times larger than the radius of the ``tube'' used in Section \ref{SimMW} to select the vertically oscillating stars in the case of an orbiting cluster. A larger selection radius may, however, be acceptable at an early stage in the spiral development. Furthermore, even with a selection radius of $1$~kpc, a broadened spiral (and therefore less pronounced ``spikes'' at the edges of the number density steps) is apparent in the example of Section \ref{SimMW}.

\begin{figure}
\centering
\includegraphics[angle=-90,width=8.0cm]{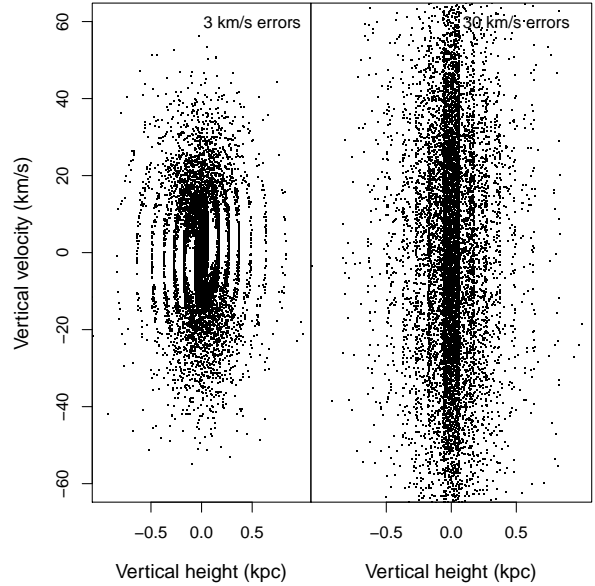}
\caption{The effect of applying random Gaussian errors with zero mean and standard deviations of $3$~km/s or $30$~km/s.}
\label{obs_errors}
\end{figure}

If we can identify the cluster stars that comprise the spiral, perhaps using the spectral type, then the main source of error in the observed spiral structure would come from the velocity errors (the position errors being negligible). Presumably in this case the subtraction of the background stars would be trivial, and we need only ensure to observe sufficiently many cluster stars to sample the phase space distribution (see Section \ref{lownstats}). Let us assume that any background contamination is removed, and, for simplicity, that the stars have entirely tangential velocities. We add Gaussian random errors to the vertical velocities of the stars in our theoretical model, where the mean of the Gaussian noise is zero and the standard deviation is either $3$~km/s or $30$~km/s. In Fig. \ref{obs_errors} we see the results at $500$~Myr after popping for a $10^6 M_{\odot}$ cluster. With Gaussian errors of $\sigma_{w} = 3$~km/s the spiral arms in the velocity direction are beginning to blur into each other, while the $\sigma_{w} = 30$~km/s errors wash out the velocity structure completely.

The situation is more difficult if we are unable to cleanly identify the cluster stars. In this case any attempt to impose a selection in radius around a suspected popped cluster (to isolate the phase space spiral) is subject to the larger distance errors discussed earlier. Furthermore, the background subtraction is no longer trivial (see Section \ref{bgndcont}).

In summary, the astrometric precision achievable by GAIA, particularly in position measurements, suggests that the spiral structure is, in principle, observable if the problems of background contamination and the vertical ``tube'' selection can be surmounted. With such high precision astrometry, and a cluster that is not too distant, it would seem that detection of the structure is feasible, at least with a cluster sample on the order of $1000$ stars.

\section{Discussion and conclusions}
\label{concl}
We have investigated the dynamical evolution of a distribution of particles that model the behaviour of a ``popping'' star cluster, in various background analytic potentials, to analyse the development of phase mixing in the phase space of the stars. Using analytical, numerical and N-body methods we have been able to elucidate the source and behaviour of the so-called ``Christmas tree'' discussed in \cite{Assmann}.

The motivating study for this work suggested that the rate of gas mass loss of the initial cluster has no significant effect on the final equilibrium state (where the gas expulsion is still taken to be acting over a short timescale in comparison to the cluster crossing time). It is nonetheless clear that a slower loss of stars from the cluster will change the evolution of the phase space spiral structure.

A slower gas expulsion rate leaves the cluster bound for longer and therefore more significantly affected by its own self-gravity. Throughout this study we have been able to neglect self-gravity as a large fraction of the initial cluster mass is lost instantaneously, leaving the Galactic potential to dominate. A very slow gas expulsion rate, on the order of tens of cluster crossing times, allows the cluster stars to adjust to the changing cluster potential, and thus will not produce a phase space spiral. Furthermore, stars that evaporate from the cluster will not do so at the same time, and so they will not populate a coherent spiral, even if there is a large number of evaporated stars. Alternatively, if the gas expulsion time is on the order of no more than a few crossing times, then the phase space spiral will still form. Gas expulsion taking place on this timescale, although far slower than the instantaneous gas loss assumed in this study, will still allow many stars to quickly escape from the cluster and populate the phase space spiral, with a small central ``bulge'' of any remaining bound stars. We have checked this with an N-body simulation, in which the gas expulsion takes place over $6$ crossing times. The influence of the cluster leads to fewer stars at high velocities, and so the spiral is less populated further out. The remaining bound cluster core modifies the dynamics of low amplitude stars, where the ``bulge'' lies, but the dynamics of escaped stars are almost entirely dictated by the background potential and so they populate the same spiral as before.

It is also worth considering the effect of a higher (or lower) star formation efficiency. In the former case the resulting cluster will be more bound, leading again to self-gravity effects that preclude the formation of the spiral for those bound stars. Any stars that escape the cluster, however, will again populate the phase space spiral as their dynamics will now be almost entirely dependent on the Galactic potential (with some small effect from the remaining bound core). A lower star formation efficiency leads again to phase mixing and spiral structure as discussed in this work, with stars that will have even more kinetic energy when the cluster pops. This is dynamically equivalent to a more massive cluster: the stars will populate the higher amplitude/velocity regions of the phase space spiral.

We summarise our main conclusions below:
\begin{itemize}
\item The so-called ``Christmas tree'' structure (i.e. stepped number density distributions in vertical position or velocity), seen in previous simulations of popping star clusters, is due to phase mixing in the vertical orbits of the stars.
\item Using a theoretical model that approximates the phase mixing phenomenon without the need for a full N-body approach, we have been able to determine various properties of the phase space spiral, such as the timescale for the spiral to wind up and the dependence of the evolution on the background potential.
\item The spiral structure is entirely specified by the analytic potential in which the cluster pops. Due to a higher velocity dispersion, higher mass clusters are able to ``populate'' higher amplitude/velocity regions of the spiral.
\item In an orbiting cluster, changing radial position leads to some particles experiencing different vertical restoring forces, and a broadening of the spiral. Those stars that move almost entirely vertically while orbiting the Galaxy (i.e. they remain at the same radius in the disk) show a clear spiral structure.
\item The main obstacle to observing the spiral is the finite resolution sampling of the phase space distribution: the structure is present for a long time, but eventually it would require an unrealistically high phase space resolution to be detected. Therefore, the presence of a higher number of stars will more fully sample the underlying phase space distribution, implying that any detection of this structure will be more likely with higher mass clusters.
\item While detection of the number density steps against a background field is likely to be extremely challenging, our chances are highest if we are lucky enough to observe a recently popped high mass cluster in a region in which the background count of high velocity stars is low. Even if the background count is high, however, the background distribution may be sufficiently smooth that it can be subtracted without leaving much noise, revealing the step.
\item Several disk heating events would ultimately wash out the structure, due to the random stellar motions spoiling the amplitude-frequency relation responsible for the spiral. One or two heating events, however, may preserve some of the structure, at least in phase space regions where the spiral has not wound up too tightly. This again suggests that we would have a better chance of observing the stepped distribution for a high mass cluster soon after popping.
\item Highly accurate observations will be required to detect the structure. The proposed positional accuracy of GAIA measurements is likely to be sufficient to detect the structure in $z$, but the velocity resolution may be insufficient for very distant clusters (i.e. beyond $\sim 10$~kpc). Added to the fact that detection is easier at an early time in the spiral development, it may be that recently popped ($< 100$~Myr ago) high mass clusters, which are not too distant, provide the best detection candidates. In this case one may find a residual bound core, surrounded by a recently formed phase space spiral of stars that have escaped the cluster.
\end{itemize}

\section*{Acknowledgments}
GC acknowledges the support of FONDECYT grant 3130480, RS acknowledges the support of FONDECYT grant 3120135, MF acknowledges the support of FONDECYT grant 1130521. PA acknowledges the support of FONDECYT grant 3130653. We thank the anonymous referee for helpful comments that improved the paper.

\appendix

\section{Series expansion of the Miyamoto-Nagai potential}
Expansion coefficients of the Miyamoto-Nagai potential for $z \ll b$.
\begin{equation}
\label{MNexpansionCoeffs}
\begin{split}
\alpha_0 &= GM_d \left( (a+b)^2 + R^2 \right)^{-1/2} \\
\alpha_1 &= GM_d \left( \frac{a}{b} + 1 \right) \left( (a+b)^2 + R^2 \right)^{-3/2} \\
\alpha_2 &= GM_d \frac{1}{2b^4}\left(a^3b + 5a^2b^2 + ab(7b^2 + R^2) + 3b^4 \right) \\
         & \times \left((a+b)^2+R^2\right)^{-5/2}
\end{split}
\end{equation}

\section{Exponential disk}
\begin{figure}
\centering
\includegraphics[angle=-90,width=8.0cm]{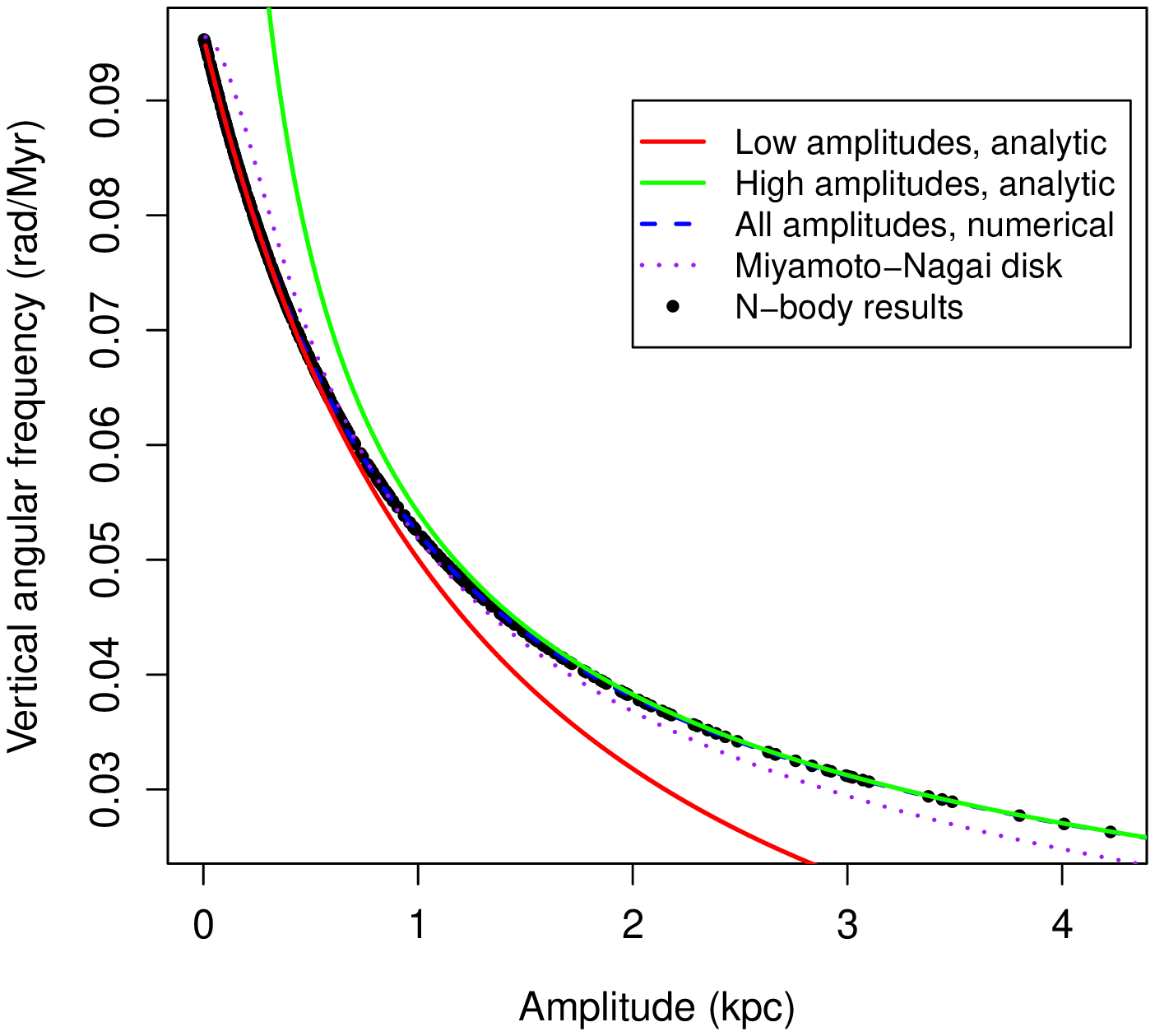}
\caption{The frequency-amplitude relation for vertical oscillations in the infinite exponential disk potential. The low amplitude analytic approximation is shown with the red solid line, while the green solid line shows the high amplitude analytic approximation. The numerical evaluation for all amplitudes is shown with the blue dashed line. For comparison, the numerically evaluated relation for the Miyamoto-Nagai disk is also shown as the purple dotted line.}
\label{expdiskfreqplot}
\end{figure}

The analytic potential of an infinite exponential disk (i.e. without a radial dependence) is given by:
\begin{equation}
\label{expdiskpot}
\Phi_{\text{Exp}} = 4\pi G\rho_0 z_0 (z_0 \exp(-|z|/z_0) + |z|),
\end{equation}
with
\begin{equation}
\rho_0 = \Sigma_0 \frac{\exp (-R/R_d) }{2z_0}.
\end{equation}
The radial scale length is set to $R_d = 3$~kpc, while the radial coordinate is set at $R=8.5$~kpc for all stars, as for the Miyamoto-Nagai disk. We choose the surface density to be $\Sigma = 1.43 \times 10^9 M_{\odot}$/kpc$^2$ to ensure that the frequency of the vertical oscillations close to the disk matches that of the Miyamoto-Nagai potential discussed in Section \ref{thesims}. This value is well within the range of realistic values for the surface density of the Galactic disk (see e.g. \citealp{BinneyTremaine}). The scale height is chosen to be $z_0 = 0.26$~kpc. Following a similar approach to that taken for the Miyamoto-Nagai disk, we can expand Eq.~\ref{expdiskpot} for small amplitude orbits, and use this approximation in Eq.~\ref{fftime} to determine the frequency dependence on amplitude for such orbits. In principle, due to the functional form of the potential, this expansion may be taken to arbitrarily high order, providing an increasingly accurate approximation to the full potential. We truncate the series expansion after the second sub-leading term to find the following approximation for the free-fall time at low amplitude:
\begin{equation}
\label{expdiskflow}
t_{ff} = \frac{1}{\sqrt{4\pi G\rho_0}} \left( \frac{\pi}{2} + \frac{A}{3z_0} + \frac{A^2}{48z_0^2} \left( \pi - \frac{8}{3} \right) \right).
\end{equation}
Similarly, we can calculate the free-fall time for high amplitude orbits by expanding the potential at large $z$ to find:
\begin{equation}
\label{expdiskfhigh}
t_{ff} = \sqrt{\frac{A}{2\pi G \rho_0 z_0}}.
\end{equation}
From these expressions, the vertical orbital frequency is easily obtained as before. The frequencies of vertical stellar orbits in an N-body simulation of a popped high mass cluster in the potential of Eq.~\ref{expdiskpot} are plotted in Fig.~\ref{expdiskfreqplot}. Included in this plot are the low and high amplitude approximations using Eqs.~\ref{expdiskflow} and \ref{expdiskfhigh} as well as the numerically evaluated frequency amplitude relation. The numerically determined relation for the Miyamoto-Nagai disk is also shown. There is little difference between the two cases, except at very high amplitudes. 

\begin{figure}
\centering
\includegraphics[width=8.0cm]{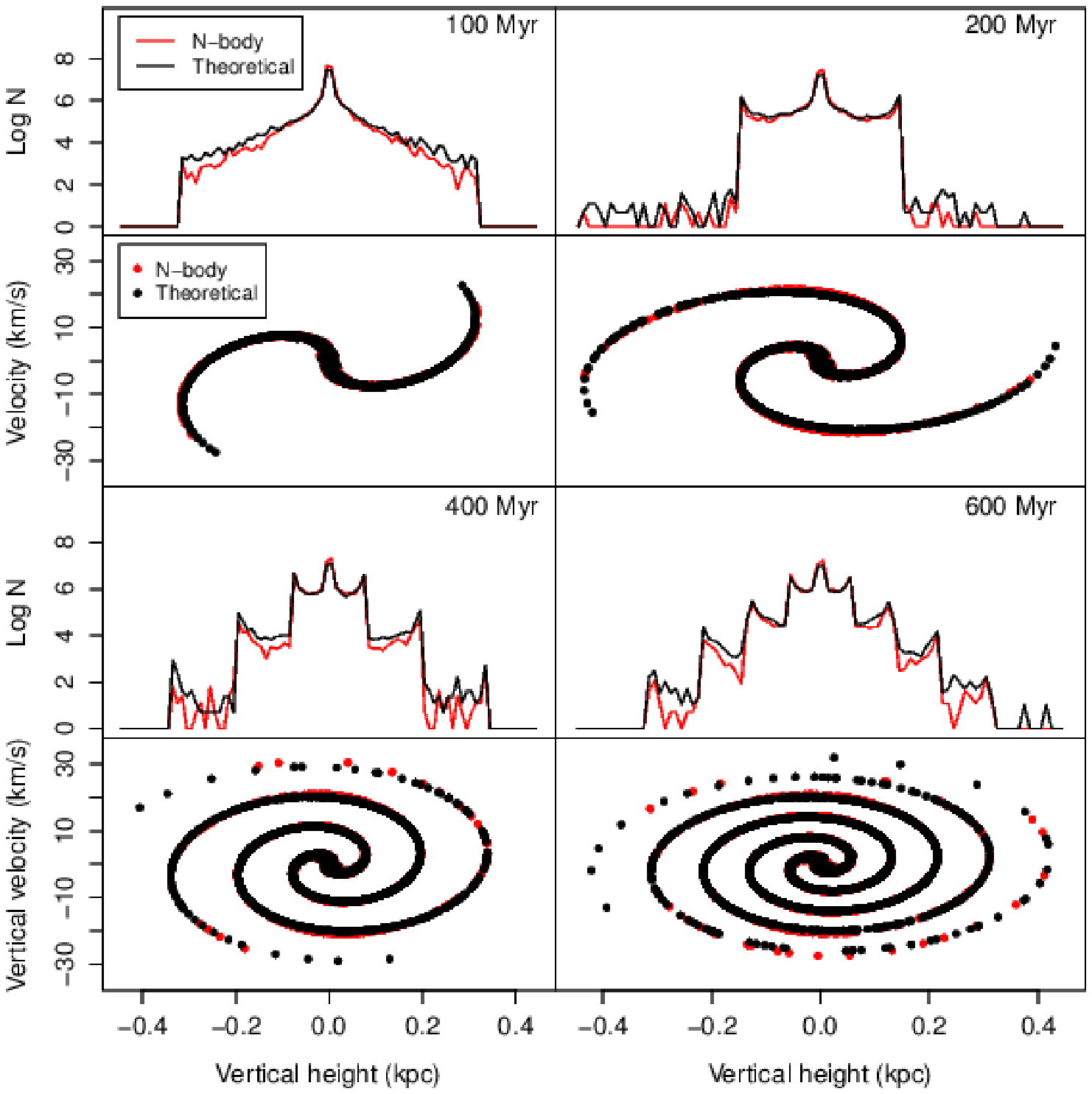}
\caption{Comparison of the theoretical model with an N-body simulation of $10^4$ stars in a $10^5 M_{\odot}$ cluster in the infinite exponential disk potential.}
\label{expdisktheorcomp}
\end{figure}

Finally, we can compare our theoretical model of a $10^5 M_{\odot}$ cluster in this potential with the N-body simulation result, as shown in Fig.~\ref{expdisktheorcomp}. As for the Miyamoto-Nagai and full Galactic potentials discussed earlier, the theoretical model agrees very well with the full N-body result.

\bsp

\label{lastpage}

\end{document}